\documentclass[lettersize,journal]{IEEEtran}
\usepackage{amsmath,amssymb,amsfonts}
\usepackage{algorithm}
\usepackage{algpseudocode}
\usepackage{array}
\usepackage[caption=false,font=normalsize,labelfont=rm,textfont=sf]{subfig}
\usepackage{textcomp}
\usepackage{stfloats}
\usepackage{url}
\usepackage{verbatim}
\usepackage{graphicx}
\usepackage{hyperref}
\usepackage{cite}
\usepackage{booktabs}
\usepackage{multirow}
\usepackage{tabularx}

\usepackage[inkscape=false]{svg}
\usepackage{siunitx}
\usepackage{makecell}
\usepackage[capitalize,nameinlink]{cleveref}

\usepackage{textgreek}   %
\usepackage{cite}
\usepackage{xcolor}
\usepackage{colortbl,booktabs}
\usepackage{bm}
\usepackage{tikz}
\usepackage{circuitikz}
\usetikzlibrary{arrows.meta,positioning,fit,backgrounds,calc,patterns.meta,decorations.pathreplacing}

\usepackage[acronym]{glossaries}
\makenoidxglossaries
\newacronym[plural={LSBs}, longplural={Least Significant Bits}]{LSB}{LSB}{Least Significant Bit}
\newacronym{MAP}{MAP}{Maximum a Posteriori Estimate}
\newacronym{iid}{i.i.d.}{independent and identically distributed}
\newacronym{GLS}{GLS}{Generalized Least Squares}
\newacronym{RUL}{RUL}{Remaining Useful Life}
\newacronym{PRBS}{PRBS}{Pseudo-Random Binary Sequence}
\newacronym{ADC}{ADC}{Analog-to-Digital Converter}
\newacronym{RK4}{RK4}{Runge-Kutta 4th Order Integrator}
\newacronym{MLE}{MLE}{Maximum Likelihood Estimate}
\newacronym{NMM}{NMM}{Noise Model Misspecification}
\newacronym{SMM}{SMM}{System Model Misspecification}
\newacronym{AR}{AR}{Autoregressive Model}
\newacronym{MCMC}{MCMC}{Markov Chain Monte Carlo}
\newacronym{ESR}{ESR}{Equivalent Series Resistance}
\newacronym{LBFGS}{LBFGS}{Limited-memory Broyden–Fletcher–Goldfarb–Shanno}
\newacronym{LM}{LM}{Levenberg--Marquardt}
\newacronym{LS}{LS}{Least Squares}
\newacronym{MSE}{MSE}{Mean Squared Error}
\newacronym{FIR}{FIR}{Finite Impulse Response}
\newacronym{IIR}{IIR}{Infinite Impulse Response}
\newacronym{PINN}{PINN}{Physics-Informed Neural Network}
\newacronym{LCR}{LCR}{Inductance-Capacitance-Resistance}
\newacronym{ANN}{ANN}{Artificial Neural Network}
\newacronym{ODE}{ODE}{Ordinary Differential Equation}
\newacronym{SNR}{SNR}{Signal-to-Noise Ratio}

\newglossaryentry{dcdcbuck}{
    name={DC-DC Buck},
    description = {A DC-DC Buck converter is a type of power supply that steps down voltage from its input (supply) to its output (load). It is a class of switched-mode power supply (SMPS) that uses a transistor as a switch to convert electrical power efficiently. The buck converter is widely used in various applications, including battery-powered devices, where it helps to extend battery life by reducing power consumption.}
}

\newglossaryentry{GPR}{
    name={GPR},
    description = {
        Gaussian Process Regression is a non-parametric regression technique that can be used to model complex systems. It works by assuming that the data points are generated from a Gaussian process, which is a collection of random variables, any finite number of which have a joint Gaussian distribution. It is then able to predict the output of the system at any point in the input space and provide a measure of uncertainty
    },
    first = {Gaussian Process Regression (GPR)}
}

\usepackage{pifont}
\usepackage{colortbl}
\definecolor{uncgray}{gray}{0.92}
\definecolor{priorbg}{HTML}{E4F4EE}   %
\definecolor{refbg}{HTML}{FBEAE0}     %
\newcolumntype{U}{>{\columncolor{uncgray}}c}  %

\usepackage{booktabs}

\definecolor{loopR}{HTML}{B23A2E}\definecolor{loopRf}{HTML}{F6E9E6}
\definecolor{blueP}{HTML}{2C6FB3}\definecolor{bluePf}{HTML}{E6EEF7}
\definecolor{ink}{HTML}{27343B}
\definecolor{edge}{HTML}{6B7780}
\definecolor{ioF}{HTML}{ECEFF1}\definecolor{ioS}{HTML}{AEBAC2}
\definecolor{fwdF}{HTML}{DCEDEA}\definecolor{fwdS}{HTML}{4E8E86}
\definecolor{nzF}{HTML}{E9E4F2}\definecolor{nzS}{HTML}{7E72A8}
\definecolor{obF}{HTML}{E4EFE3}\definecolor{obS}{HTML}{6E9B6C}
\definecolor{amF}{HTML}{F7E8C7}\definecolor{amS}{HTML}{B98A2E}

\newcommand{\thetahat}{\hat{\bm{\theta}}}
\newcommand{\R}{\mathbb{R}}
\newcommand{\N}{\mathcal{N}}
\newcommand{\D}{\mathcal{D}}
\newcommand{\Loss}{\mathcal{L}}

\newsavebox{\panelA}\newsavebox{\panelB}\newsavebox{\panelC}

\begin{document}

\title{Uncertainty-Aware Parameter Estimation for Condition Monitoring of Power Converters}

\author{Tomas Monopoli, Jiahong Liu,~\IEEEmembership{Member,~IEEE}, and Shuai Zhao,~\IEEEmembership{Senior Member,~IEEE}%
\thanks{\noindent This work is supported by the Phy-Caliper project sponsored by the Villum Foundation. (corresponding author: Shuai Zhao)}%
\thanks{
\noindent T. Monopoli, J.~Liu and S.~Zhao are with AAU Energy, Aalborg University, 9220 Aalborg, Denmark. (e-mail: tmon@energy.aau.dk; jiahongliu@ieee.org; szh@energy.aau.dk;).}}%
\markboth{Submitted to Transactions on Power Electronics}%
{Monopoli \MakeLowercase{\textit{et al.}}: Uncertainty-Aware Parameter Estimation for Power Converters}

\maketitle

\begin{abstract}
Changes in circuit parameters can indicate component degradation in power converters, but measurement noise and limited parameter observability can make these changes difficult to detect reliably. This paper proposes an uncertainty-aware parameter estimation framework to distinguish genuine parameter changes from variations caused by estimation uncertainty. Using a differentiable switched converter model, the method requires only one or two samples per switching period. Model-derived residual whitening and a Bayesian posterior approximation quantify estimation uncertainty and reveal poorly identifiable parameters. Simulations demonstrate improved estimation accuracy over least-squares and physics-informed neural-network benchmarks, with uncertainty estimates consistent with repeated noise realizations. Experiments on a laboratory buck converter validate the estimated parameters against independently derived references and demonstrate the detection of a controlled output-capacitance reduction. The proposed framework provides a basis for reliable parameter-change detection in power converter condition monitoring.
\end{abstract}

\begin{IEEEkeywords}
Bayesian parameter estimation, buck DC--DC converters, uncertainty quantification, condition monitoring.
\end{IEEEkeywords}

\section{Introduction}
Power converters are widely employed to improve energy efficiency and enable advanced functionalities in systems such as renewable energy installations, electric vehicles,
aerospace platforms, and industrial drives \cite{VanderBroeck2023}. The increasingly central role of these systems makes their reliability and operational safety a critical
concern \cite{Wang2021}, generating significant research interest. A key approach to improving converter reliability is to actively monitor the converter state of health before critical failures occur. The state of health of a converter is often characterized by the values of its circuit parameters, which can deviate from their nominal values due to manufacturing tolerances, environmental conditions, and component degradation \cite{Miao2024,dc_link_monitoring2}. Tracking vulnerable components such as electrolytic capacitors and power transistors can identify soft faults before they evolve into hard failures \cite{Miao2024,dc_link_monitoring2,AlGreer2019,Lu2023,Shuai2022}. Furthermore, the  trajectories of parameter estimates can be used to predict the remaining useful life of equipment and proactively schedule maintenance \cite{Mann2016,Lu2023,DiNezio2023,DiNezio2022}. 

In the literature, the estimation of circuit parameters from measurement data is commonly referred to as \emph{parameter identification}, and it has been the subject of extensive research in the power electronics community. \Cref{tab:method_landscape} maps the main approaches. Traditional parameter identification methods inject broadband excitation signals, most commonly a \gls{PRBS}, into the control loop to elicit informative responses from the system \cite{AlGreer2019,AlGreer2012}. However, such approaches can degrade converter performance and are impractical in real-world applications \cite{Miao2024,Song2024}. As a result, recent research has focused on non-intrusive methods that estimate parameters using data collected under transient operating conditions \cite{AlGreer2019,dc_link_monitoring,Nazif2025}. 

As \cref{tab:method_landscape} shows, these methods can be classified based on whether they require a training dataset or the dynamic equations of the converter. Learning-based methods obtain the mapping from measurements to parameters with an artificial neural network \cite{soliman2017artificial,lin2019non,Kalker2022,oruklu2024machine,She2022,Shang2024}, trained on a large dataset that is typically generated by simulating the converter under varying parameter values. Since they do not rely on the dynamic equations, they are the alternative when no adequate model exists, or when a high-fidelity model is too costly to run but a dataset can be obtained. However, the trained models are sensitive to any difference between the training simulations and the real operating conditions, which can lead to poor estimation performance in practice. Within this family, \glspl{PINN} add some dynamic equations to the training objective of the neural network, which enables gradient-based optimization and improves the consistency with the system dynamics \cite{Dey2025}.

Physics-driven methods, to which this paper belongs, rely instead on the dynamic equations of the converter, and they fit the parameters to the board's own measurements, without a training dataset. The methods of this family differ in how they carry out the fit. Mathematical modeling evaluates the circuit equations directly on operating data. For example, \cite{wechsler2012condition} monitors the DC-link capacitors of a drive from the voltage and current sensors already in place; however, such an estimate is specific to one component and is not a joint estimate of the circuit parameters. Metaheuristic optimization algorithms, such as particle swarm optimization \cite{Nazif2025,Song2024,DiNezio2022,DiNezio2023} and genetic algorithms \cite{de2023real}, fit a complete digital twin of the converter to the observed data. They require many evaluations of the simulator, which is computationally expensive, and offer no guarantee of convergence to a global optimum. The \gls{PINN}-based method in~\cite{Shuai2022} introduces a gradient-based approach which embeds the dynamic equations in the training loss and learns the circuit parameters together with the network on the converter's own measurements. Its network training nevertheless remains slow, and its least-squares loss gives only a point estimate. The method developed in this paper uses a gradient-based optimizer as well and fits the dynamic equations directly on a differentiable forward model, without a neural network.

\begin{table*}[!t]
    \caption{Parameter identification in power electronic converters: two main families, methods, principles, main limitations and properties.}
    \label{tab:method_landscape}
    \centering
    {\definecolor{ink}{HTML}{1F1F1F}\definecolor{mute}{HTML}{666666}\definecolor{rule}{HTML}{BDBDBD}%
\definecolor{cp}{HTML}{0072B2}\definecolor{cpd}{HTML}{00507F}%
\definecolor{bphys}{HTML}{E8F0F8}\definecolor{blearn}{HTML}{F3EEE8}%
\definecolor{glearn}{HTML}{8A4B24}%
\def\lsm{\fontsize{6.3}{7.3}\selectfont}%
\def\lbd{\fontsize{7}{8.2}\selectfont}%
\begin{tikzpicture}[x=1cm, y=1cm, every node/.style={font=\lsm, text=ink},
    txt/.style={anchor=north west, align=left, inner xsep=0pt, inner ysep=0pt,
                text width=#1, execute at begin node={\hyphenpenalty=10000}},
    sub/.style={anchor=west, inner xsep=0pt, inner ysep=0pt, font=\lsm\bfseries\scshape},
    head/.style={anchor=south, align=center, inner sep=1pt, font=\lsm\bfseries},
    headl/.style={anchor=south west, align=left, inner xsep=0pt, inner ysep=1pt,
                  font=\lsm\bfseries},
    gl/.style={line width=0.5pt},
    pics/full/.style={code={\fill[ink] (0,0) circle (2.1pt);}},
    pics/half/.style={code={\draw[ink, gl] (0,0) circle (1.9pt);
                            \fill[ink] (0,1.9pt) arc (90:270:1.9pt) -- cycle;}},
    pics/none/.style={code={\draw[ink, gl] (0,0) circle (1.9pt);}},
    pics/na/.style={code={\draw[mute, line width=0.5pt] (-2pt,0) -- (2pt,0);}}]

\def\xl{0}\def\xr{18.1}
\def\xu{0}\def\xv{1.2}          %
\def\xt{1.5}                    %
\def\xa{1.62}\def\wa{3.45cm}     %
\def\xb{5.15}\def\wb{4.5cm}      %
\def\xc{9.75}\def\wc{4.3cm}      %
\def\cx{{14.62,15.62,16.62,17.62}}
\def\rh{0.80}                    %
\def\sh{0.32}                    %
\def\bh{0.62}                    %
\newcommand{\mrow}[6]{%
  \node[txt=\wa] at (\xa,#1-0.1) {{\lbd #2}\\[0.6pt]{\color{mute}#3}};
  \node[txt=\wb] at (\xb,#1-0.1) {#4};
  \node[txt=\wc] at (\xc,#1-0.1) {#5};
  \foreach \g [count=\k from 0] in {#6} {\pgfmathsetmacro{\xx}{\cx[\k]}%
    \pic at (\xx,#1-0.5*\rh) {\g};}}
\newcommand{\hair}[1]{\draw[rule!70, line width=0.25pt] (\xt,#1) -- (\xr,#1);}
\newcommand{\block}[5]{%
  \fill[#2] (\xt,#1) rectangle (\xr,#1-\bh);
  \node[anchor=north west, inner xsep=0pt, inner ysep=0pt, font=\lbd\bfseries, text=#3]
       at (\xa,#1-0.07) {#4};
  \node[anchor=north west, inner xsep=0pt, inner ysep=0pt, text width=16.2cm, align=left]
       at (\xa,#1-0.33) {#5};}
\newcommand{\inputtab}[4]{%
  \pgfmathsetmacro{\ymid}{0.5*(#1+#2)}%
  \fill[#3] (\xu,#1-0.04) -- (\xv,#1-0.04) -- (\xv,\ymid+0.3) -- (\xt-0.02,\ymid)
            -- (\xv,\ymid-0.3) -- (\xv,#2+0.04) -- (\xu,#2+0.04) -- cycle;
  \node[align=center, text width=1.15cm, inner sep=0pt] at ({0.5*(\xu+\xv)},\ymid) {#4};}

\node[headl] at (\xu,0.04) {Input\strut};
\node[headl] at (\xa,0.04) {Method\strut};
\node[headl] at (\xb,0.04) {Principle\strut};
\node[headl] at (\xc,0.04) {Main limitation\strut};
\node[head] at (14.62,0.04) {non-\\intrusive\strut};
\node[head] at (15.62,0.04) {low\\cost\strut};
\node[head] at (16.62,0.04) {noise\\model\strut};
\node[head] at (17.62,0.04) {param.\\uncert.\strut};
\draw[ink, line width=0.6pt] (\xl,0.04) -- (\xr,0.04);

\pgfmathsetmacro{\ya}{-0.06}
\block{\ya}{blearn}{glearn}{Learning-based}
  {The mapping from measurements to parameters is learned from a training dataset; suited when
   no adequate model exists, or one is too costly to run.}
\pgfmathsetmacro{\yb}{\ya-\bh}
\mrow{\yb}{Neural networks}{\cite{soliman2017artificial, lin2019non, Kalker2022,
    oruklu2024machine, She2022, Shang2024}}
  {Learn the mapping from measured waveforms to parameters, trained on simulated data.}
  {Large labeled datasets; poor extrapolation across operating points; not interpretable.}
  {full, half, none, none}
\pgfmathsetmacro{\yc}{\yb-\rh}\hair{\yc}
\mrow{\yc}{Physics-informed ML}{\cite{Xie2025, Dey2025, Lu2023}}
  {Embeds the circuit differential equations in the training loss of a neural network.}
  {Delicate training (loss weighting, stiff switching dynamics); high offline cost.}
  {full, none, none, none}
\pgfmathsetmacro{\ye}{\yc-\rh}
\draw[ink, line width=0.6pt] (\xt,\ye) -- (\xr,\ye);

\pgfmathsetmacro{\yg}{\ye-0.2}
\draw[ink, line width=0.6pt] (\xt,\yg) -- (\xr,\yg);
\block{\yg}{bphys}{cpd}{Physics-driven}
  {The parameters of the converter's dynamic equations are fitted to its own measurements; suited
   when the equations describe the board closely enough.}
\pgfmathsetmacro{\yh}{\yg-\bh}
\mrow{\yh}{Signal injection}{\cite{heo2013capacitance, pirsto2020real, li2021online,
    yao2023estimation, Ahmeid2018, Ahmeid2017}}
  {Injects test signals (PRBS, frequency response) and fits a circuit or transfer-function
   model to the response.}
  {Perturbs operation and power quality; needs extra hardware; limited to scheduled tests.}
  {none, na, none, none}
\pgfmathsetmacro{\yi}{\yh-\rh}\hair{\yi}
\mrow{\yi}{Mathematical modeling}{\cite{wechsler2012condition}}
  {Evaluates the circuit equations, e.g.\ of the DC-link capacitor, on operating data.}
  {Specific to one component; no joint estimate of the circuit parameters.}
  {full, full, none, none}
\pgfmathsetmacro{\yj}{\yi-\rh}\hair{\yj}
\mrow{\yj}{Metaheuristic optimization}{\cite{peng2020digital, liu2021condition,
    diNezio2023parameters, fard2023digital, de2023real, XLi2016, Nazif2025, Song2024,
    DiNezio2022, DiNezio2023}}
  {Particle swarm, genetic or other evolutionary algorithms fit a digital twin of the converter to its
   measurements.}
  {Many simulator evaluations; no convergence guarantee; sensitive to tuning.}
  {full, none, none, none}
\pgfmathsetmacro{\yk}{\yj-\rh}\hair{\yk}
\mrow{\yk}{Probabilistic methods}{\cite{milton2020controller, zimmer2026parameter}}
  {Polynomial chaos or Gaussian processes on a model of the converter.}
  {High computational cost; the uncertainty is that of the signals, and the parameters are point
   estimates.}
  {full, none, full, none}
\pgfmathsetmacro{\yp}{\yk-\rh}\hair{\yp}
\mrow{\yp}{Physics-informed network}{\cite{Shuai2022}}
  {Embeds the dynamic equations in the loss of a neural network trained on the converter's own
   measurements.}
  {Slow network training; unweighted least-squares loss.}
  {full, none, none, none}
\pgfmathsetmacro{\yn}{\yp-\rh}\hair{\yn}
\begin{scope}[on background layer]
  \fill[cp!5] (\xt,\yn) rectangle (\xr,\yn-\rh);
\end{scope}
\mrow{\yn}{\textbf{This work}}{Sections~\ref{sec:methodology}--\ref{sec:val_hardware}}
  {Whitened likelihood of the circuit model, one sample per dwell; Laplace posterior.}
  {Needs the switching instants; weakly identified parameters ($R_\mathrm{ds,on}$, $V_F$)
   keep wide intervals.}
  {full, full, full, full}
\pgfmathsetmacro{\ym}{\yn-\rh}
\draw[ink, line width=0.6pt] (\xl,\ym) -- (\xr,\ym);

\inputtab{\ya}{\ye}{blearn}{{\lbd\bfseries\color{glearn}training\\dataset}}
\inputtab{\yg}{\ym}{bphys}{{\lbd\bfseries\color{cpd}dynamic\\equations}}

\node[anchor=north east, text=mute, inner xsep=0pt] at (\xr,\ym-0.1)
    {\tikz[baseline=-0.6ex]\fill (0,0) circle (2.1pt);~yes\enspace
     \tikz[baseline=-0.6ex]{\draw[line width=0.5pt] (0,0) circle (1.9pt);
       \fill (0,1.9pt) arc (90:270:1.9pt) -- cycle;}~partly\enspace
     \tikz[baseline=-0.6ex]\draw[line width=0.5pt] (0,0) circle (1.9pt);~no\enspace
     \tikz[baseline=-0.6ex]\draw[line width=0.5pt] (-2pt,0) -- (2pt,0);~not reported.};
\end{tikzpicture}%
}
\end{table*}

Probabilistic tools, such as polynomial chaos expansions and Gaussian processes, have also been applied to models of the converter \cite{milton2020controller,zimmer2026parameter}. They describe the uncertainty of the signals, but none return a validated uncertainty on the estimated parameters. To the authors' knowledge, all methods in \cref{tab:method_landscape} therefore produce \emph{point estimates} of the parameters, without quantifying the uncertainty associated with them. This is a significant gap, since a point estimate by itself gives no indication of estimation quality, and estimation reliability can vary widely depending on noise level, data quantity and data quality. Furthermore, when used for condition-monitoring applications it cannot directly distinguish whether a parameter drifts because of the noise inherent in the estimation process or due to a genuine change. Both limitations can be overcome by returning a \emph{distribution} over the parameters rather than a single value. 

The objective of this work is to estimate the parameters of a power converter together with a posterior distribution whose intervals match the actual scatter of the estimates. The estimator needs only one sample per dwell, i.e.\ per ON or OFF interval of the switch, or one sample per switching period. These are the samples that a digital controller takes, and they lie away from the switching transitions, where the switched model is least accurate. The parameters are those of the switched dynamic equations of the converter, the loss is the negative log-likelihood of its measurements, and the shape of that loss is the posterior. The contributions are:
\begin{enumerate}
    \item A \emph{Bayesian formulation of the estimation}: parameter estimation is recast from the minimization of a loss into the solution of a probabilistic problem, the inference of a posterior distribution over the parameters. Every estimate then carries a credible interval, estimates can be compared with each other and with a reference in probabilistic terms, and prior knowledge, such as the datasheet value and tolerance of a component, enters as a prior distribution.
    \item A \emph{noise model derived from the converter model}: each measured sample is used twice, as the end of one prediction step and as the start of the next, so the prediction errors of neighboring steps share the same measurement noise and are not independent. In order to obtain correct uncertainty quantification, the correlation between consecutive residuals is taken into account and removed by whitening filters derived from the dynamic equations.
    \item A \emph{validation of accuracy and uncertainty}: on simulated data, the estimator is more accurate than least squares, increasingly so as the noise grows. Its uncertainty is validated first in simulation, where the stated intervals are compared with the scatter of the estimates over repeated noisy datasets, and then on a laboratory buck converter, against independent component references and repeated records.
    \item An \emph{application to condition monitoring}: on the laboratory converter, the posteriors of successive records detect and flag a change of the output capacitor, while the other parameters are found constant.
\end{enumerate}

\cref{sec:theory} establishes the statistical framework. \cref{sec:methodology} describes the
converter model, the sampling strategies, the likelihood and its whitening, the
optimizer, and the complete pipeline. \cref{sec:val_simulated} validates the
estimator on simulated data with a known ground truth, and \cref{sec:val_hardware}
on a laboratory buck converter. \cref{sec:conclusion} presents the conclusions.

\section{Statistical Framework}
\label{sec:theory}

This section sets up the statistical framework of the estimator. The loss minimized is the negative log-likelihood of the data, completed by the prior (\cref{subsec:bayesian_setup}); minimizing it selects the parameters under which the observed data are most probable (\cref{subsec:loss_meaning}), and the shape of this loss is the posterior, which carries the uncertainty of the estimate (\cref{subsec:loss_posterior}). \Cref{subsec:posterior_methods} reviews how the posterior can be obtained.

\subsection{Bayesian Parameter Estimation}
\label{subsec:bayesian_setup}

Let $\D$ denote the observed dataset and $\bm{\theta} \in \R^P$ the vector of unknown parameters, where $P$ is the number of parameters. The Bayesian framework treats $\bm{\theta}$ as a random variable and summarizes all information about it through the posterior distribution, obtained from Bayes' rule:
\begin{equation}
    p(\bm{\theta} \mid \D) = \frac{p(\D \mid \bm{\theta})\, p(\bm{\theta})}{p(\D)},
    \label{eq:bayes_rule}
\end{equation}
where $p(\D \mid \bm{\theta})$ is the likelihood, $p(\bm{\theta})$ the prior, and $p(\D)$ the evidence. Because the evidence does not depend on $\bm{\theta}$, the most probable parameter vector given the data is obtained by maximizing the numerator alone, or equivalently by minimizing the negative log-posterior:
\begin{equation}
    \thetahat
    = \arg\min_{\bm{\theta}}\; [
    \underbrace{-\log p(\D \mid \bm{\theta})}_{\text{data term}}
    \;+\;
    \underbrace{\bigl(-\log p(\bm{\theta})\bigr)}_{\text{prior term}}
    ].
    \label{eq:map_loss}
\end{equation}
The data term is the negative log-likelihood of the measurements under the model; its form for the residuals of the converter model is derived in \cref{subsec:loss_function}. This is the loss minimized throughout the paper.

\subsection{The Loss as a Negative Log-Probability}
\label{subsec:loss_meaning}

The loss~\eqref{eq:map_loss} is not an arbitrary measure of the mismatch between model and measurements: it is a negative log-probability, and this is what gives its minimum and its shape a statistical meaning. The model maps a parameter vector $\bm{\theta}$ to a prediction of the measurements, and the measurements are assumed to differ from this prediction by random noise of known distribution. Each $\bm{\theta}$ therefore defines a probability distribution $p(\D \mid \bm{\theta})$ of the data the experiment would produce, and the likelihood evaluates this distribution at the data actually observed. Parameters that move the prediction away from the data leave a systematic deviation that such noise would rarely produce, and their likelihood falls accordingly. The \gls{MLE} is the parameter vector that maximizes the likelihood, i.e.\ that minimizes the data term $-\log p(\D \mid \bm{\theta})$ of~\eqref{eq:map_loss} alone, and under which the observed data are therefore most probable. This choice is also theoretically optimal since, for identifiable parameters, the \gls{MLE} is asymptotically unbiased and efficient, i.e.\ its covariance attains the Cram\'er--Rao bound, the smallest covariance that any unbiased estimator using the same data can have~\cite[Ch.~8]{vanderVaart1998}.

The prior term plays a similar role, where $-\log p(\bm{\theta})$ penalizes parameter values that are improbable before the data are seen. The minimizer $\thetahat$ of the complete loss~\eqref{eq:map_loss}, with both terms, is the \gls{MAP}, which weighs the two sources of information against each other. In power electronics such knowledge is readily available: a datasheet gives the nominal value of a component and its tolerance. Where the data determine a parameter well, the data term dominates and the prior has little effect; where they do not, the prior keeps the estimate within the range the datasheet allows. As the data come to dominate the prior, the \gls{MAP} estimate inherits the optimality of the \gls{MLE}. 

\subsection{The Shape of the Loss Is the Posterior}
\label{subsec:loss_posterior}

Let $\Loss(\bm{\theta}) = -\log p(\D \mid \bm{\theta}) - \log p(\bm{\theta})$ denote the loss of~\eqref{eq:map_loss}. By Bayes' rule~\eqref{eq:bayes_rule},
\begin{equation}
    p(\bm{\theta} \mid \D) \;\propto\; \exp\!\bigl[-\Loss(\bm{\theta})\bigr],
    \label{eq:posterior_shape}
\end{equation}
so the posterior is the shape of the loss. Its minimum is the estimate $\thetahat$, and how steeply the loss rises around it is how well the data determine the parameters: a narrow valley means a well-determined parameter, a wide or flat one a poorly determined one.

This reading holds only if $\Loss$ is the negative log-likelihood of a correctly specified model: the noise distribution, including its correlation from sample to sample, must be the one that generated the data. A loss that is not, such as a least-squares loss on correlated residuals, still has a minimum and a shape, but that shape need not match the uncertainty, and the intervals it implies can be too wide or too narrow. For a correctly specified model and informative data, the width of the posterior also matches the scatter of the estimate over repeated experiments~\cite[Ch.~10]{vanderVaart1998}. This is tested in \cref{sec:val_simulated} over repeated noise realizations, and the likelihood of the converter model is built to meet the condition (\cref{subsec:loss_function,subsec:whitening}).

\subsection{Obtaining the Posterior}
\label{subsec:posterior_methods}

\begin{figure*}[!tp]
    \centering
    \includesvg[inkscapearea=page]{figures/generated/fig_posterior_profiles}
    \caption{Loss of one simulated load-step record over two parameter pairs, the other
    parameters re-optimized at every point, against its Laplace approximation at three noise levels, given by the \acrfull{SNR} of the current / voltage, with the noisiest case as a surface on the right (one sample per dwell):
    (a)~$L$ and $C$; (b)~$V_F$ and $R_\mathrm{ds,on}$. Axes in units of each parameter's Laplace standard deviation~$\sigma$.}
    \label{fig:posterior_profiles}
\end{figure*}
For a nonlinear forward model the shape~\eqref{eq:posterior_shape} has no closed form. Monte Carlo sampling, e.g.~Markov chain Monte Carlo, characterizes it without approximation, but needs an evaluation of the loss for each of many thousands of samples. The cheaper route is a local approximation, which replaces the loss by a quadratic at its minimum.

At its minimum the gradient of $\Loss$ vanishes, and a second-order Taylor expansion from \eqref{eq:posterior_shape} gives
\begin{equation}
    \Loss(\bm{\theta}) \approx \Loss(\thetahat)
    + \tfrac{1}{2}(\bm{\theta} - \thetahat)^\top \mathbf{H}\, (\bm{\theta} - \thetahat),
    \label{eq:taylor_expansion}
\end{equation}
where $\mathbf{H}$ is the Hessian of the loss,
\begin{equation}
    \mathbf{H} = \nabla_{\bm{\theta}}^2 \Loss(\bm{\theta})\big|_{\bm{\theta} = \thetahat}.
    \label{eq:hessian_def}
\end{equation}
Substituting this quadratic into the posterior~\eqref{eq:posterior_shape}, the constant factor $\exp[-\Loss(\thetahat)]$ is absorbed by the normalization, and what remains is the exponential of a quadratic form,
\begin{equation}
    p(\bm{\theta} \mid \D) \;\propto\;
    \exp\!\bigl[-\tfrac{1}{2}(\bm{\theta} - \thetahat)^\top \mathbf{H}\, (\bm{\theta} - \thetahat)\bigr],
    \label{eq:laplace_posterior}
\end{equation}
which is the density of a Gaussian $\N(\thetahat, \mathbf{H}^{-1})$ with mean $\thetahat$ and covariance $\mathbf{H}^{-1}$. This is the \textit{Laplace approximation} of the posterior~\cite{Murphy2022}, which only requires the curvature of the loss function at its minimum to be constructed. With this approximation, high curvature along a direction means a small posterior variance along it, low curvature a large one, and the off-diagonal entries of $\mathbf{H}^{-1}$ give the correlations between parameters.

\Cref{fig:posterior_profiles} compares the loss with its local quadratic approximation over two pairs of parameters of the converter model of \cref{sec:methodology}, for three records that differ only in their measurement noise. The noise of each record is expressed by the \gls{SNR} of each measured signal,
\begin{equation}
    \mathrm{SNR} = 20 \log_{10} \frac{x_\mathrm{rms}}{\sigma_x}~\text{dB},
    \label{eq:snr}
\end{equation}
where $x_\mathrm{rms}$ is the RMS value of the switching ripple of the signal $x$ ($i_L$ or $v_o$) and $\sigma_x$ the standard deviation of the noise added to it. The three records have an \gls{SNR} of 40, 26 and \qty{20}{dB} on the current and of 16, 2 and \qty{-4}{dB} on the voltage. These levels, which range from low to high noise, are further utilized and motivated in \cref{sec:val_simulated}.
Even in the noisiest record, whose voltage noise exceeds its ripple, some parameter subspaces remain well described by the quadratic (\cref{fig:posterior_profiles}a), while others, more weakly identified, depart from it (\cref{fig:posterior_profiles}b). This is the price of the approximation: it is fast, but it is only a local description of the posterior.

\section{Methodology}
\label{sec:methodology}
This section describes the estimation framework. \cref{subsec:buck_model}
introduces the buck converter and its forward model, \cref{subsec:sampling} the sampling strategies, and \cref{subsec:non_intrusive} why they need nothing beyond what the converter already measures. \cref{subsec:loss_function} derives the likelihood of the
one-step residuals and \cref{subsec:whitening} the whitening operator that makes it
tractable. \cref{subsec:optimizer} describes the optimizer, and \cref{subsec:overall_pipeline}
assembles the pieces into the complete pipeline, from the record to the Laplace posterior.

\subsection{Buck Converter Model}
\label{subsec:buck_model}
\begin{figure}[t]
    \centering
    {\definecolor{ink}{HTML}{1F1F1F}\definecolor{mute}{HTML}{8A8A8A}%
\definecolor{meas}{HTML}{0072B2}%
\def\bcsz{\fontsize{7}{8.2}\selectfont}%
\def\bchd{\fontsize{8.5}{10}\selectfont}     %
\newcommand{\bcmeas}[1]{\textcolor{meas}{$#1$}}%
\begin{circuitikz}[line width=0.45pt, color=ink, font=\bcsz,
    every node/.append style={font=\bcsz, text=ink},
    title/.style={font=\bchd, anchor=base west, inner sep=0pt},
    eqbox/.style={anchor=north, fill=ink!5, rounded corners=1.5pt, inner xsep=4pt,
                  inner ysep=2.5pt},
    port/.style={anchor=center, text=meas},
    iloop/.style={meas, line width=0.5pt, dash pattern=on 1.8pt off 1.2pt,
                  rounded corners=2.5pt, -{Stealth[length=4pt, width=3.4pt]}}]
\ctikzset{voltage=american, bipoles/length=0.8cm, bipoles/thickness=1.6,
          resistors/scale=0.7, inductors/scale=0.72, capacitors/scale=0.72,
          diodes/scale=0.62, batteries/scale=0.72, switches/scale=0.75,
          current/distance=0.5, bipole current style/.style={color=meas},
          bipole label style/.style={text=ink}}

\node[title] at (-0.75, 2.5) {\textbf{(a)}\enspace Buck converter};
\draw (0,0) to[battery1, l=$V_\mathrm{in}$] (0,1.9)
      to[cosw, l=$S$] (1.1,1.9)
      to[R, l=$R_\mathrm{ds,on}$] (2.25,1.9) -- (2.45,1.9)
      to[cute inductor, l=$L$, i>_=\bcmeas{i_L}] (3.6,1.9)
      to[R, l=$R_L$] (4.75,1.9) -- (6.8,1.9);
\draw (2.45,0) to[D*, l=$V_F$, *-*] (2.45,1.9);
\draw (5.3,1.9) to[R, l_=$R_C$, *-] (5.3,0.95) to[C, l_=$C$] (5.3,0);
\draw (6.15,1.9) to[R, l=$R$, *-*] (6.15,0);
\draw (0,0) -- (6.8,0);
\draw (5.3,0) node[circ]{};
\draw (6.8,1.9) node[ocirc]{} (6.8,0) node[ocirc]{};
\node[port] at (6.8,1.62) {$+$};
\node[port] at (6.8,0.95) {$v_o$};
\node[port] at (6.8,0.28) {$-$};

\begin{scope}[shift={(0,-2.75)}]
\node[title] at (-0.75, 2.03) {\textbf{(b)}\enspace Switch on, $S = 1$};
\draw (0,0) to[battery1, l=$V_\mathrm{in}$] (0,1.5)
      to[R, l=$R_\mathrm{ds,on}$] (1.0,1.5)
      to[cute inductor, l=$L$] (2.0,1.5)
      to[R, l=$R_L$] (3.0,1.5) -- (3.25,1.5)
      to[generic] (3.25,0) -- (0,0);
\node[port] at (3.6,1.22) {$+$};
\node[port] at (3.6,0.75) {$v_o$};
\node[port] at (3.6,0.28) {$-$};
\draw[iloop] (0.42,0.62) |- (2.85,1.12) |- (0.42,0.38) -- (0.42,0.56);
\node[text=meas] at (1.63,0.75) {$i_L$};
\node[eqbox] at (1.55,-0.18) {$\begin{aligned}
    L\,\frac{\mathrm{d}i_L}{\mathrm{d}t} &= V_\mathrm{in} - v_o\\[-1pt]
    &\quad - (R_\mathrm{ds,on}+R_L)\,i_L
  \end{aligned}$};
\end{scope}

\begin{scope}[shift={(4.55,-2.75)}]
\node[title] at (-0.75, 2.03) {\textbf{(c)}\enspace Switch off, $S = 0$};
\draw (0,0) to[D*, l=$V_F$] (0,1.5) -- (0.25,1.5)
      to[cute inductor, l=$L$] (1.25,1.5)
      to[R, l=$R_L$] (2.25,1.5) -- (2.5,1.5)
      to[generic] (2.5,0) -- (0,0);
\node[port] at (2.85,1.22) {$+$};
\node[port] at (2.85,0.75) {$v_o$};
\node[port] at (2.85,0.28) {$-$};
\draw[iloop] (0.42,0.62) |- (2.1,1.12) |- (0.42,0.38) -- (0.42,0.56);
\node[text=meas] at (1.26,0.75) {$i_L$};
\node[eqbox] at (1.2,-0.18) {$\begin{aligned}
    L\,\frac{\mathrm{d}i_L}{\mathrm{d}t} &= -V_F - v_o\\[-1pt]
    &\quad - R_L\,i_L
  \end{aligned}$};
\end{scope}
\end{circuitikz}%
}
    \caption{Buck converter model: (a)~circuit, whose labeled elements (all but the switch $S$)
    are the unknown parameters, with the measured signals in blue; (b)~switch on, $S = 1$;
    (c)~switch off, $S = 0$.}
    \label{fig:buck_topology}
\end{figure}

The \gls{dcdcbuck} converter is a fundamental building block of power-electronic systems, and the degradation of its components has been extensively studied in the literature~\cite{dc_link_monitoring,dc_link_monitoring2}. The unknown parameters $\bm{\theta}$ are the labeled elements of \cref{fig:buck_topology}a: the inductance $L$ and its series resistance $R_L$, the output capacitance $C$ and its \gls{ESR} $R_C$, the MOSFET on-resistance $R_\mathrm{ds,on}$, the load $R$, the input voltage $V_\mathrm{in}$ and the diode forward drop $V_F$. The measured signals are the inductor current $i_L$ and the output voltage $v_o$. With the state $\mathbf{x} = [i_L,\, v_o]^\top$ and the switch state $S \in \{0,1\}$ (1 for ON, 0 for OFF), Kirchhoff's laws give
\begin{align}
    \frac{di_L}{dt}
    &= -\frac{S\,R_\mathrm{ds,on} + R_L}{L}\, i_L
    \;-\; \frac{1}{L}\, v_o
    \;+\; \frac{S\,V_\mathrm{in} - (1-S)\,V_F}{L}, \notag \\[2pt]
    \frac{dv_o}{dt}
    &= \frac{R_C\,R}{R_C+R}\,\frac{di_L}{dt}
    + \frac{R}{C\,(R_C+R)}\, i_L
    - \frac{1}{C\,(R_C+R)}\, v_o.
    \label{eq:buck_dynamics}
\end{align}
The two cases of the inductor equation correspond to the two current loops of \cref{fig:buck_topology}b,c.

Within each regime, \eqref{eq:buck_dynamics} is affine in the state,
\begin{equation}
    \dot{\mathbf{x}} = \mathbf{A}_S(\bm{\theta})\,\mathbf{x} + \mathbf{b}_S(\bm{\theta}),
    \label{eq:affine_dynamics}
\end{equation}
with $\mathbf{A}_S$ and $\mathbf{b}_S$ read off~\eqref{eq:buck_dynamics} for $S = 0$ or $1$. The forward model integrates it with the classical explicit \gls{RK4} scheme~\cite{Hairer1993}, where for $\dot{\mathbf{x}} = f(\mathbf{x})$ and a step of length $h$,
\begin{equation}
\begin{gathered}
    \mathbf{k}_1 = f(\mathbf{x}_n), \quad
    \mathbf{k}_2 = f\bigl(\mathbf{x}_n + \tfrac{h}{2}\mathbf{k}_1\bigr), \\
    \mathbf{k}_3 = f\bigl(\mathbf{x}_n + \tfrac{h}{2}\mathbf{k}_2\bigr), \quad
    \mathbf{k}_4 = f\bigl(\mathbf{x}_n + h\,\mathbf{k}_3\bigr), \\
    \hat{\mathbf{x}}_{n+1} = \mathbf{x}_n + \tfrac{h}{6}\bigl(\mathbf{k}_1 + 2\mathbf{k}_2 + 2\mathbf{k}_3 + \mathbf{k}_4\bigr).
\end{gathered}
\label{eq:rk4}
\end{equation}
Since $f$ is affine, every stage $\mathbf{k}_i$ is affine in $\mathbf{x}_n$, and so is the step since substituting~\eqref{eq:affine_dynamics} into~\eqref{eq:rk4} gives the one-step predictor
\begin{equation}
    \hat{\mathbf{x}}_{n+1} = F(\mathbf{x}_n,\, S_n,\, \Delta t_n;\, \bm{\theta}) = \mathbf{J}_n(\bm{\theta})\, \mathbf{x}_n + \mathbf{c}_n(\bm{\theta}),
    \label{eq:forward_predictor}
\end{equation}
\begin{equation}
    \mathbf{J}_n = \sum_{k=0}^{4} \frac{(h\mathbf{A}_S)^k}{k!}, \qquad
    \mathbf{c}_n = h \sum_{k=0}^{3} \frac{(h\mathbf{A}_S)^k}{(k+1)!}\, \mathbf{b}_S,
    \label{eq:rk4_affine}
\end{equation}
with $S = S_n$ and $h = \Delta t_n$. Here $F$ is the \gls{RK4} forward model over the interval from sample $n$ to sample $n+1$, and $S_n$ and $\Delta t_n$ are the switch state and the length of that interval. $\mathbf{J}_n$ and $\mathbf{c}_n$ are explicit polynomials in the entries of $\mathbf{A}_S$ and $\mathbf{b}_S$, so the whole predictor is differentiable with respect to $\bm{\theta}$, which enables gradient-based optimization through automatic differentiation. How an interval is integrated depends on where the samples are taken, which is the subject of the next subsection.

\subsection{Sampling Strategies}
\label{subsec:sampling}
\begin{figure}[t]
    \centering
    {\input{figures/generated/fig_sampling_strategies.tex}}
    \caption{Switching-point sampling (orange), one sample per dwell (blue) and one sample per period (green) over two switching periods (illustrative): waveforms, samples and
    integration steps (top); for each scheme, the residual of one interval (bottom left) and
    the common form of the step, one affine map of the state per interval (bottom right).}
    \label{fig:sampling_strategies}
\end{figure}
The predictor~\eqref{eq:forward_predictor} advances the state over the interval between two
samples, and requires the switch state, and hence the dynamics~\eqref{eq:buck_dynamics}, to be constant over each step. Three sampling strategies, switching-point sampling, one sample per dwell and one sample per period, meet this requirement in different ways (\cref{fig:sampling_strategies}).

\paragraph*{Switching-point sampling} One sample is taken at each switching instant (orange),
as in~\cite{Shuai2022}. Every interval then lies within one regime and is integrated as one RK4
step with $S_n$ held constant. Consecutive samples sit on the turning points of the ripple, so
their difference is the full current excursion of the dwell, a direct measurement of the
inductor slope.

However, a real converter exhibits non-ideal switching behavior, where ringing, finite switching times and the delays of the measurement chain make the switched model biased in practice. This is illustrated in \cref{fig:sampling_strategies}, where the real converter (red) rings through every switching transition, while the ideal switched model (black) does not describe it.

\paragraph*{One sample per dwell} A dwell is the interval between two consecutive switching instants, over which the switch is either ON or OFF. One sample is taken at the center of each dwell (blue), and \emph{one sample per dwell} refers to this placement throughout the paper. The switching instant $t_s$ inside each interval is recorded with the samples (\cref{subsec:non_intrusive}), and the interval is integrated as two RK4 sub-steps split at $t_s$, of lengths $\Delta t_{n,1} = t_s - t_n$ and
$\Delta t_{n,2} = t_{n+1} - t_s$, each with the switch state of its own dwell held constant.

\paragraph*{One sample per period} This strategy extends the previous one to fewer samples. A digital controller that samples once per switching period takes the center of one of the two dwells (\cref{subsec:non_intrusive}), i.e.\ every other sample of the previous strategy. Each interval then spans a whole period and contains two switching instants, and is integrated as three RK4 sub-steps, the first and the last with the switch state of the sampled dwell and the middle one with the other. The same construction holds for any number $K$ of switching instants inside an interval, which is then integrated as $K+1$ sub-steps. The samples can therefore, in principle, be spaced further apart than one switching period, although this is not tested here.

\paragraph*{A common form} A composition of affine sub-steps is itself an affine map; for the two sub-steps of one sample per dwell,
\begin{equation}
    \mathbf{J}_n = \mathbf{J}_{n,2}\,\mathbf{J}_{n,1}, \qquad
    \mathbf{c}_n = \mathbf{J}_{n,2}\,\mathbf{c}_{n,1} + \mathbf{c}_{n,2},
    \label{eq:centre_point_step}
\end{equation}
and a third sub-step composes in the same way, see \cref{fig:sampling_strategies}. Under every strategy the predictor therefore has the same form~\eqref{eq:forward_predictor} and is differentiable in $\bm{\theta}$, so everything that follows applies equally to all of them.

The price of moving the samples away from the edges is information: the ripple largely cancels
between two dwell centers, so consecutive samples follow the slower envelope of the waveform.
The parameters of the inductor equation ($L$, $R_L$, $R_\mathrm{ds,on}$, $V_\mathrm{in}$, $V_F$)
are therefore less directly observed than under switching-point sampling. It is nevertheless the preferred option on real converters, whose switching edges the idealized model~\eqref{eq:buck_dynamics} does not describe.
\subsection{Non-Intrusive Measurement}
\label{subsec:non_intrusive}
The inductor current $i_L$ and the output voltage $v_o$ are the signals a converter measures for its closed-loop control. Therefore, the estimator is applied to these same measurements to see the performance it can achieve in practice without requiring any extra sensors.

The sampling strategies of \cref{subsec:sampling} are also chosen to coincide with the ways a digital controller samples these signals. Its \gls{ADC} is triggered synchronously with the PWM carrier, and two placements of the samples are common~\cite{Buso2015}.

\paragraph*{Mean-value sampling} The sample is taken at a peak or a valley of the carrier, which corresponds to the center of a dwell (\cref{fig:sampling_strategies}). A controller that updates its duty cycle once per period samples once per period, at the center of one dwell. This is the conventional case~\cite{corradini2008multisampled} and the acquired samples correspond exactly to the one-sample-per-period strategy (green in \cref{fig:sampling_strategies}), which the estimator can therefore reuse as they are. At high switching frequencies, a controller may even sample less often than once per period~\cite{lukic2005sigma}, a case not tested here but to which this estimator naturally extends. Sampling at both carrier extremes, as a controller that updates twice per period does, gives one sample per dwell (blue).

\paragraph*{Peak and valley sampling} The samples are taken at the switching instants, where the inductor current reaches the minimum and the maximum of its ripple. Sampling at both gives the switching-point samples (orange), also two per period and within the reach of the \gls{ADC}, but inside the switching transitions.

\paragraph*{Sampling and switching instants} The switching instants are the crossings of the carrier with the modulating signal (\cref{fig:sampling_strategies}), but can equally be read off the gate signal or computed from the duty cycle. The switching instants are therefore available from the controller for the strategies with one sample per dwell or per period, which require this information.

\subsection{Loss Function}
\label{subsec:loss_function}
From the observed state at sample $n$, the predictor~\eqref{eq:forward_predictor} gives the next state, and its mismatch with the observed next state $\mathbf{x}_{n+1}^{\mathrm{obs}}$ is the residual
\begin{equation}
\label{eq:residual_definition}
\mathbf{r}_n(\bm{\theta}) = F(\mathbf{x}_n^{\mathrm{obs}}, S_n, \Delta t_n;\, \bm{\theta}) - \mathbf{x}_{n+1}^{\mathrm{obs}},
\end{equation}
stacked over the record as $\mathbf{r} = (\mathbf{r}_0^\top, \ldots, \mathbf{r}_{N-1}^\top)^\top$. Each $\bm{\theta}$ turns the record into its own sequence of residuals, and the likelihood of the data under $\bm{\theta}$ is the probability the residuals would follow if $\bm{\theta}$ were the true parameters. Choosing the loss of \cref{sec:theory} is therefore choosing the distribution of the residuals.

The measured states are corrupted by additive white noise
$\bm{\varepsilon}_n \sim \mathcal{N}(\mathbf{0}, \bm{\Sigma}_x)$, so that
$\mathbf{x}_n^{\mathrm{obs}} = \mathbf{x}_n + \bm{\varepsilon}_n$. When the
parameters equal their true values $\bm{\theta} = \bm{\theta}^\star$, a
first-order Taylor expansion of the residual around the true state gives
\begin{align}
  \label{eq:residual_first_order}
  \mathbf{r}_n(\bm{\theta}^\star)
    &\approx
      \underbrace{F(\mathbf{x}_n, S_n, \Delta t_n;\,
                    \bm{\theta}^\star)
      - \mathbf{x}_{n+1}}_{=\,0}
      +\; \mathbf{J}_n\bm{\varepsilon}_n - \bm{\varepsilon}_{n+1} \notag \\
    &= \mathbf{J}_n\bm{\varepsilon}_n - \bm{\varepsilon}_{n+1},
\end{align}
where $\mathbf{J}_n = \partial F / \partial \mathbf{x}\big|_{\mathbf{x}_n}$ is the
Jacobian of the forward model, available analytically
from~\eqref{eq:forward_predictor}. Since this is a linear function of Gaussian
random variables, the per-step residual covariance follows directly from the
measurement noise $\bm{\Sigma}_x$:
\begin{equation}
    \bm{\Sigma}_{r,n} = \bm{\Gamma}_{0,n} = \bm{\Sigma}_x + \mathbf{J}_n\,\bm{\Sigma}_x\,\mathbf{J}_n^\top.
    \label{eq:gamma0_method}
\end{equation}
Even though the measurement noise is white, the
residuals are \emph{not} sequentially independent, because each residual
$\mathbf{r}_n$ shares the noise term $\bm{\varepsilon}_n$ with its neighbor
$\mathbf{r}_{n-1}$ through the predictor input. The lag covariances follow
directly from~\eqref{eq:residual_first_order}:
\begin{align}
    \bm{\Gamma}_{1,n} &= \mathrm{Cov}(\mathbf{r}_n,\, \mathbf{r}_{n-1}) \notag \\
     &= \mathrm{Cov}(\mathbf{J}_n\,\bm{\varepsilon}_n - \bm{\varepsilon}_{n+1},\, \mathbf{J}_{n-1}\,\bm{\varepsilon}_{n-1} - \bm{\varepsilon}_n) \notag \\
     &= -\,\mathbf{J}_n\,\bm{\Sigma}_x, \label{eq:gamma1_method}\\[2pt]
    \bm{\Gamma}_{k,n} &= 0 \qquad \text{for } |k| \geq 2. \label{eq:gammak_method}
\end{align}
The residual sequence therefore has a vector moving-average structure of order
one, and the corresponding covariance matrix $\bm{\Sigma}_r$ is
block-tridiagonal,
\begin{equation}
  \bm{\Sigma}_r =
  \begin{bmatrix}
    \bm{\Gamma}_{0,0} & \bm{\Gamma}_{1,1}^\top & & & \\[2pt]
    \bm{\Gamma}_{1,1} & \bm{\Gamma}_{0,1} & \bm{\Gamma}_{1,2}^\top & & \\[2pt]
     & \bm{\Gamma}_{1,2} & \ddots & \ddots & \\[2pt]
     & & \ddots & \bm{\Gamma}_{0,N-2} & \bm{\Gamma}_{1,N-1}^\top \\[2pt]
     & & & \bm{\Gamma}_{1,N-1} & \bm{\Gamma}_{0,N-1}
  \end{bmatrix},
  \label{eq:sigma_r_structure}
\end{equation}
with $\bm{\Gamma}_{0,n}$ from~\eqref{eq:gamma0_method} on the main diagonal,
$\bm{\Gamma}_{1,n}$ from~\eqref{eq:gamma1_method} on the first sub-diagonal, and
all remaining blocks zero by~\eqref{eq:gammak_method}.
Therefore, the residuals distribution implied by the model at the true parameters is Gaussian with zero mean and this block-tridiagonal covariance. The negative log-likelihood of the data is therefore
\begin{equation}
    \Loss_\mathrm{data}(\bm{\theta}) = -\log p(\D \mid \bm{\theta})
    = \tfrac{1}{2}\, \mathbf{r}(\bm{\theta})^\top \bm{\Sigma}_r^{-1}\, \mathbf{r}(\bm{\theta}) + \text{const.},
    \label{eq:loss_general}
\end{equation}
Together with the prior, this is the loss~\eqref{eq:map_loss} the optimizer minimizes. The inverse of $\bm{\Sigma}_r$ is dense and computationally expensive, but the quadratic form can be evaluated without forming it, as described next.

\subsection{Model-Derived Whitening}
\label{subsec:whitening}
We seek an operator $\mathbf{T}$ with $\bm{\Sigma}_r^{-1} = \mathbf{T}^\top \mathbf{T}$, so that the quadratic form in~\eqref{eq:loss_general} becomes the squared norm of the \emph{whitened} residuals,
\begin{equation}
    \mathbf{r}^\top \bm{\Sigma}_r^{-1}\, \mathbf{r} \;=\; \|\mathbf{T} \mathbf{r}\|^2 = \sum_n \|\mathbf{w}_n\|^2,
    \label{eq:whitened_quadratic}
\end{equation}
where the innovations $\mathbf{w}_n$ are uncorrelated and of unit variance.

The covariances $\bm{\Gamma}_{0,n}$ and $\bm{\Gamma}_{1,n}$ change along the record with the switching regime and the operating point. The whitener replaces them by their averages over the record, $\bm{\Gamma}_0$ and $\bm{\Gamma}_1$. For this block-tridiagonal and block-Toeplitz $\bm{\Sigma}_r$, $\mathbf{T}$ follows from its block LDL factorization~\cite{Golub2013}, which reduces to the recursion
\begin{equation}
\begin{aligned}
    \mathbf{D}_0 &= \bm{\Gamma}_0, &
    \mathbf{K}_k &= \bm{\Gamma}_1 \mathbf{D}_{k-1}^{-1}, &
    \mathbf{D}_k &= \bm{\Gamma}_0 - \mathbf{K}_k \bm{\Gamma}_1^\top, \\
    \dot{\mathbf{r}}_0 &= \mathbf{r}_0, &
    \dot{\mathbf{r}}_k &= \mathbf{r}_k - \mathbf{K}_k \dot{\mathbf{r}}_{k-1}, &
    \mathbf{w}_k &= \mathbf{L}_k^{-1} \dot{\mathbf{r}}_k,
\end{aligned}
\label{eq:ldl_recursion}
\end{equation}
with $\mathbf{L}_k \mathbf{L}_k^\top = \mathbf{D}_k$. Expanding $\dot{\mathbf{r}}_k$ in the residuals $\mathbf{r}_k, \mathbf{r}_{k-1}, \ldots, \mathbf{r}_0$ gives a whitening filter whose taps depend on the position $k$ in the record and whose length grows with it. In this form the whitener is a sequential recursion over the whole record, rebuilt from $\mathbf{J}_n(\bm{\theta})$ whenever $\bm{\theta}$ changes, which is slow to evaluate and to differentiate inside the optimizer.

At position $m$, the taps of this filter are exactly the solution $\bm{\Phi}_1, \ldots, \bm{\Phi}_m$ of the order-$m$ block Yule--Walker equations~\cite{Lutkepohl2005},
\begin{equation}
    \sum_{j=1}^{m} \bm{\Phi}_j\, \bm{\Gamma}_{k-j} = \bm{\Gamma}_k, \qquad k = 1, \ldots, m,
    \qquad \bm{\Gamma}_{-k} = \bm{\Gamma}_k^\top,
    \label{eq:yule_walker}
\end{equation}
and $\mathbf{D}_m$ is their innovation covariance. We therefore use the exact whitener with its length frozen at $m$,
\begin{equation}
    \mathbf{w}_n = \mathbf{L}_{\bar n}^{-1} \Bigl( \mathbf{r}_n - \sum_{k=1}^{\bar n} \bm{\Phi}^{(\bar n)}_k\, \mathbf{r}_{n-k} \Bigr),
    \qquad \bar n = \min(n, m),
    \label{eq:ramp_whitener}
\end{equation}
where $\bm{\Phi}^{(\bar n)}_k$ are the taps at position $\bar n$, read off~\eqref{eq:ldl_recursion} after $\bar n$ steps. The first $m$ innovations are those of the factorization; beyond them, \eqref{eq:ramp_whitener} is a fixed matrix FIR filter of order $m$, applied to the whole record as one convolution in $\mathcal{O}(N m)$ operations. For a scalar residual that is the plain difference of two consecutive noise samples, i.e.\ for a Jacobian equal to one, the taps $\phi_k$ fall linearly with the lag $k$, $\phi_k = -(m+1-k)/(m+1)$, hence the name \emph{ramp whitener}. \Cref{fig:whitening} shows the whitener built at the estimate of one simulated record. \Cref{fig:whitening_acf} gives the correlation of the one-step residuals with their own past, for each channel and between the two: before whitening it is significant at lag one only, at the value predicted by~\eqref{eq:gamma0_method}--\eqref{eq:gamma1_method}, and after whitening it stays within the 95\,\% bounds of white noise. \Cref{fig:whitening_taps} shows the taps of the order-120 whitener, and \cref{fig:whitening_filter} shows how~\eqref{eq:ramp_whitener} is applied: each innovation $\mathbf{w}_n$ is the current residual minus the past ones weighted by $\bm{\Phi}_k$ and scaled by $\mathbf{L}^{-1}$, while at the start of the record the window holds only the residuals so far.
\begin{figure}[t]
    \centering
    \subfloat[]{\includesvg[inkscapearea=page]{figures/generated/fig_whitening_acf}\label{fig:whitening_acf}}\hfill
    \subfloat[]{\includesvg[inkscapearea=page]{figures/generated/fig_whitening_taps}\label{fig:whitening_taps}}\\[2pt]
    \subfloat[]{{\definecolor{ink}{HTML}{1F1F1F}\definecolor{mute}{HTML}{6E6E6E}%
\definecolor{cp}{HTML}{0072B2}\definecolor{cpd}{HTML}{00507F}%
\def\wsm{\fontsize{6.3}{7.2}\selectfont}%
\def\wbd{\fontsize{7}{8.2}\selectfont}%
\begin{tikzpicture}[every node/.style={font=\wbd, text=ink}, >={Stealth[length=3pt,width=2.6pt]},
  res/.style={draw=ink!60, fill=ink!6, rounded corners=0.8pt, minimum width=0.4cm,
              minimum height=0.34cm, inner sep=0pt, line width=0.4pt},
  sum/.style={circle, draw=ink, fill=white, inner sep=0pt, minimum size=0.3cm, line width=0.5pt},
  inn/.style={draw=cpd, fill=cp!14, rounded corners=0.8pt, minimum width=0.4cm,
              minimum height=0.34cm, inner sep=0pt, line width=0.5pt},
  tap/.style={font=\wsm, fill=white, inner sep=0.6pt},
  lab/.style={font=\wsm, text=mute}]
\node[res] (h0) at (0.30,0.0) {};
\node[lab, below=1.2pt] at (h0.south) {$\mathbf{r}_0$};
\node[res] (h1) at (0.95,0.0) {};
\node[lab, below=1.2pt] at (h1.south) {$\mathbf{r}_1$};
\node[res] (h2) at (1.60,0.0) {};
\node[lab, below=1.2pt] at (h2.south) {$\mathbf{r}_2$};
\node[sum] (hs) at (1.60,1.15) {\tiny$+$};
\node[inn] (hw) at (1.60,1.85) {};
\node[font=\wsm, anchor=south] at (hw.north) {$\mathbf{w}_2$};
\draw[->, ink, line width=0.6pt] (h2) -- (hs);
\draw[->, ink!49, line width=0.72pt, shorten >=0.4pt] (h0.north) -- (hs);
\draw[->, ink!70, line width=1.12pt, shorten >=0.4pt] (h1.north) -- (hs);
\draw[->, cpd, line width=0.6pt] (hs) -- node[right, font=\wsm, inner sep=1.5pt] {$\mathbf{L}_2^{-1}$} (hw);
\draw[decorate, decoration={brace, mirror, amplitude=3pt}, mute, line width=0.4pt] (0.06,-0.50) -- (1.84,-0.50) node[midway, below=2.5pt, font=\wsm, text=mute, align=center] {record start ($n < m$):\\the exact taps of position $n$};
\node[res] (s0) at (2.95,0.0) {};
\node[lab, below=1.2pt] at (s0.south) {$\mathbf{r}_{n-m}$};
\node[font=\wbd, text=mute] (s1) at (3.65,0.0) {$\cdots$};
\node[res] (s2) at (4.40,0.0) {};
\node[lab, below=1.2pt] at (s2.south) {$\mathbf{r}_{n-4}$};
\node[res] (s3) at (5.12,0.0) {};
\node[lab, below=1.2pt] at (s3.south) {$\mathbf{r}_{n-3}$};
\node[res] (s4) at (5.84,0.0) {};
\node[lab, below=1.2pt] at (s4.south) {$\mathbf{r}_{n-2}$};
\node[res] (s5) at (6.56,0.0) {};
\node[lab, below=1.2pt] at (s5.south) {$\mathbf{r}_{n-1}$};
\node[res] (s6) at (7.28,0.0) {};
\node[lab, below=1.2pt] at (s6.south) {$\mathbf{r}_n$};
\node[sum] (ss) at (7.28,1.15) {\tiny$+$};
\node[inn] (sw) at (7.28,1.85) {};
\node[font=\wsm, anchor=south] at (sw.north) {$\mathbf{w}_n$};
\draw[->, ink, line width=0.6pt] (s6) -- node[tap, pos=0.3] {$\mathbf{I}$} (ss);
\draw[->, ink!28, line width=0.30pt, shorten >=0.4pt] (s0.north) -- node[tap, pos=0.22, text=ink!55] {$-\bm{\Phi}_m$} (ss);
\draw[->, ink!87, line width=1.45pt, shorten >=0.4pt] (s2.north) -- (ss);
\draw[->, ink!88, line width=1.46pt, shorten >=0.4pt] (s3.north) -- (ss);
\draw[->, ink!88, line width=1.48pt, shorten >=0.4pt] (s4.north) -- node[tap, pos=0.46, text=ink!88] {$-\bm{\Phi}_2$} (ss);
\draw[->, ink!90, line width=1.50pt, shorten >=0.4pt] (s5.north) -- node[tap, pos=0.2, text=ink!90] {$-\bm{\Phi}_1$} (ss);
\draw[->, cpd, line width=0.6pt] (ss) -- node[right, font=\wsm, inner sep=1.5pt] {$\mathbf{L}_m^{-1}$} (sw);
\draw[->, mute, line width=0.4pt, dash pattern=on 1.2pt off 1pt] (7.60,1.85) -- (7.90,1.85) node[right, font=\wsm, text=mute, inner sep=1pt] {$n{+}1$};
\draw[decorate, decoration={brace, mirror, amplitude=3pt}, mute, line width=0.4pt] (2.71,-0.50) -- (7.52,-0.50) node[midway, below=2.5pt, font=\wsm, text=mute, align=center] {steady state ($n \geq m$): a fixed filter of order $m$,\\slid one sample per innovation};
\node[font=\wsm, text=mute] at (2.28,0.00) {$\cdots$};
\end{tikzpicture}%
}\label{fig:whitening_filter}}
    \caption{The model-derived whitener on one simulated record (load steps, one sample per dwell, \gls{SNR} of \qty{20}{dB} on the current and \qty{-4}{dB} on the voltage): (a)~correlation of the residuals with their own past, before and
    after whitening; (b)~taps $\bm{\Phi}_k$ of the order-120 whitener; (c)~application of the
    whitener along the record.}
    \label{fig:whitening}
\end{figure}
\begin{figure*}[!tp]
    \centering
    \includegraphics[width=0.9\textwidth]{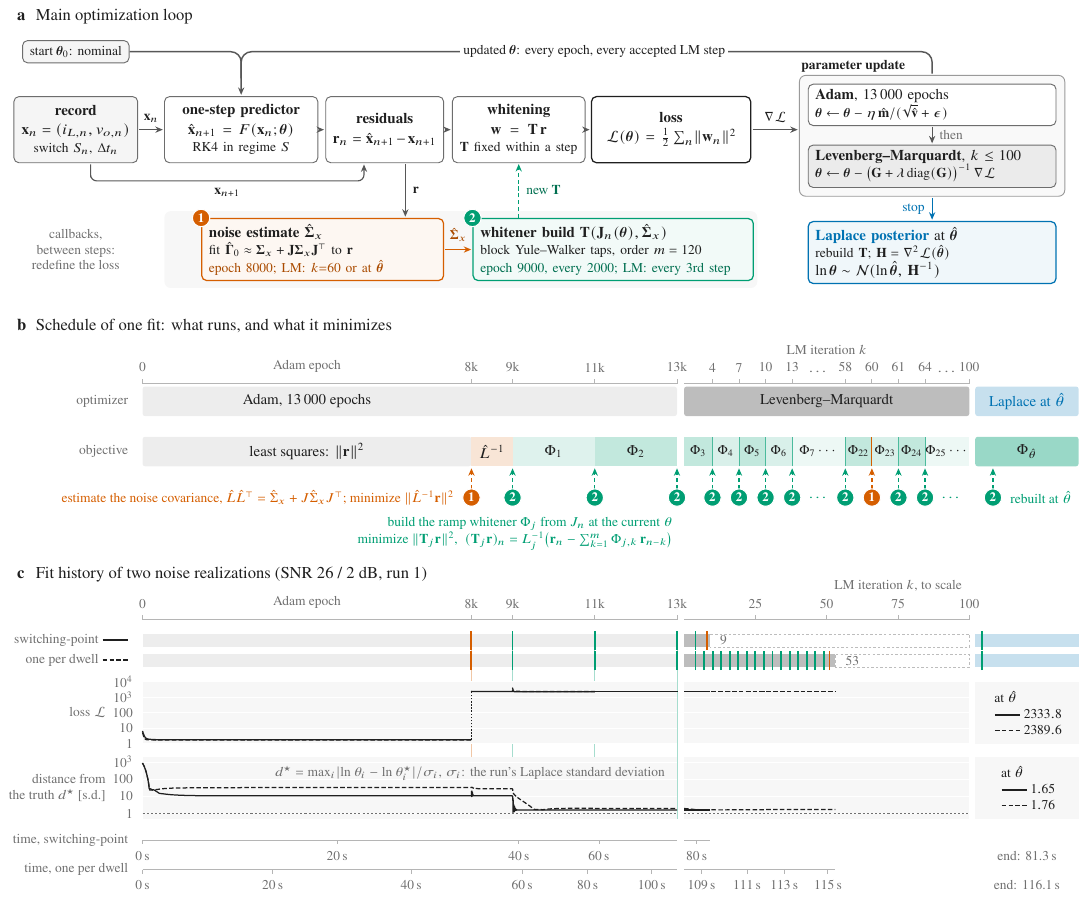}
    \caption{Estimation pipeline: (a)~main optimization loop and its two callbacks;
    (b)~schedule of one fit; (c)~history of two fits of one simulated record (\gls{SNR} of \qty{26}{dB} on the current and \qty{2}{dB} on the voltage),
    one per sampling scheme; the loss function can increase abruptly after a callback, since each callback redefines the loss function.}
    \label{fig:single_loop}
\end{figure*}

The order $m$ follows from the model as well. The recursion~\eqref{eq:ldl_recursion} is the innovations algorithm of the moving-average residual~\cite{Brockwell1991}. Its gain converges to a steady state $\mathbf{K}_\infty$, and the taps of the whitener decay with the lag as $\rho^k$, with $\rho$ the spectral radius of $\mathbf{K}_\infty$. The order is set to the lag at which the taps have fallen to 3\,\%, $m = \lceil \ln 0.03 / \ln \rho \rceil$: about 120 on the simulated records and 10 to 16 on the hardware records.
\subsection{Optimizer}
\label{subsec:optimizer}
The loss~\eqref{eq:map_loss}, with the data term~\eqref{eq:loss_general} evaluated through
the whitener~\eqref{eq:ramp_whitener}, is differentiable in $\bm{\theta}$ through the RK4
predictor. It is minimized in two phases, both acting on $\log\bm{\theta}$ so that a step
changes every parameter by a relative amount: Adam first, then \gls{LM}.

Adam~\cite{Kingma2015} scales each component of the gradient by running averages of the
gradient and of its square,
\begin{equation}
    \bm{\theta}_{t+1} = \bm{\theta}_t - \eta\, \frac{\hat{\mathbf{m}}_t}{\sqrt{\hat{\mathbf{v}}_t} + \epsilon},
    \label{eq:adam_step}
\end{equation}
where $\hat{\mathbf{m}}_t$ and $\hat{\mathbf{v}}_t$ are the bias-corrected exponential moving averages of the gradient
and of its elementwise square, $\eta$ is the learning rate and $\epsilon$ a small constant.
Each parameter moves by at most about $\eta$ per step, whatever the scale of its gradient, which
makes Adam robust from the nominal starting point, far from the optimum, where the gradients of
different parameters span orders of magnitude and a quadratic model of the loss is poor. Being diagonal, however, it cannot follow directions that couple several parameters, and with a
fixed learning rate it does not settle exactly on the optimum.

\gls{LM} then refines the estimate, once Adam has brought $\bm{\theta}$ into the basin where
its quadratic model holds. The data term is a sum of squares of the whitened residuals
$\mathbf{w}(\bm{\theta})$ of~\eqref{eq:whitened_quadratic}, so its Hessian is approximated by the
Gauss--Newton matrix $\mathbf{G} = \mathbf{J}_w^\top \mathbf{J}_w$~\cite{Nocedal2006}, with
$\mathbf{J}_w = \partial \mathbf{w} / \partial \bm{\theta}$ computed by forward-mode automatic
differentiation through the predictor and the whitener. \gls{LM} steps with
this matrix, damped,
\begin{equation}
    \bm{\theta}_{t+1} = \bm{\theta}_t - \bigl(\mathbf{G}_t + \lambda_t \operatorname{diag}(\mathbf{G}_t)\bigr)^{-1} \nabla_{\bm{\theta}} \Loss(\bm{\theta}_t),
    \label{eq:lm_step}
\end{equation}
where the damping by $\operatorname{diag}(\mathbf{G}_t)$ makes the step invariant to the scale
of each parameter~\cite{Marquardt1963}: a large $\lambda_t$ gives a short, scaled gradient step, a
small one the Gauss--Newton step. Because it uses the full Gauss--Newton curvature at every
step, \gls{LM} remains effective when the Hessian is ill-conditioned and the parameter
sensitivities differ by orders of magnitude~\cite{Nocedal2006}, as is common in
parameter estimation~\cite{Transtrum2010,Transtrum2011}. Quasi-Newton methods such as
the limited-memory BFGS method~\cite{Liu1989,Nocedal2006} instead approximate the curvature from a few recent
steps, and can stall along the weakly determined directions of such problems.

\subsection{Overall Optimization Approach}
\label{subsec:overall_pipeline}

The optimizer described above is wrapped in an estimation pipeline. \cref{fig:single_loop}a shows its main loop and \cref{fig:single_loop}b when each of its steps acts. One iteration predicts every sample from the previous one with~\eqref{eq:forward_predictor}, whitens the residuals with the latest $\mathbf{T}$~\eqref{eq:ramp_whitener}, and updates $\bm{\theta}$ from the gradient of the loss. The whitener and the noise model are regularly updated using callbacks during the optimization. In particular, callback~\textcircled{\scriptsize 1} estimates the noise covariance $\bm{\Sigma}_x$ from the residuals, and callback~\textcircled{\scriptsize 2} rebuilds $\mathbf{T}$ from $\mathbf{J}_n(\bm{\theta})$ and $\bm{\Sigma}_x$. \Cref{fig:single_loop}b shows the optimizer in use over one fit, with every callback at the iteration where it fires. \Cref{fig:single_loop}c follows two fits of the same simulated record, one for the switching-point sampling and one for the one-per-dwell sampling. It shows the callbacks as they fired, the loss, the wall time, and the distance of the worst parameter from the truth in units of its Laplace standard deviation. The pipeline executes the following stages.
\begin{enumerate}
    \item \textbf{Initialization.} Set $\bm{\theta}_0$ to the nominal values from datasheet specifications. The noise covariance starts at a placeholder, $\bm{\Sigma}_x = \mathbf{I}$, and the residuals are weighted by the static Cholesky factor of~\eqref{eq:gamma0_method} at $\bm{\theta}_0$.
    \item \textbf{Adam on the provisional loss.} Adam~\eqref{eq:adam_step} brings $\bm{\theta}$ into the basin of the \gls{MAP}. The placeholder noise makes this loss a different objective from the likelihood, and its minimum lies far from the final estimate (\cref{fig:single_loop}c), but it provides the residuals from which the noise is estimated.
    \item \textbf{Noise estimate} (callback~\textcircled{\scriptsize 1}). If $\bm{\Sigma}_x$ is not known a priori, it is fitted to the empirical residual covariance through $\hat{\bm{\Gamma}}_0 \approx \bm{\Sigma}_x + \mathbf{J}_n \bm{\Sigma}_x \mathbf{J}_n^\top$, from~\eqref{eq:gamma0_method}.
    \item \textbf{Whitener} (callback~\textcircled{\scriptsize 2}). The ramp whitener~\eqref{eq:ramp_whitener} is built from $\mathbf{J}_n(\bm{\theta})$ at the current $\bm{\theta}$ and from $\hat{\bm{\Sigma}}_x$, and held fixed until the next build (\cref{subsec:whitening}). 
    \item \textbf{Levenberg--Marquardt.} \gls{LM}~\eqref{eq:lm_step} refines $\bm{\theta}$; a step is kept only if it lowers the loss. Between accepted steps the whitener is rebuilt every $n$ iterations (callback~\textcircled{\scriptsize 2}), and the noise can be re-estimated (callback~\textcircled{\scriptsize 1}).
    \item \textbf{Laplace posterior.} The whitener is rebuilt at $\thetahat$ and $\mathbf{H} = \nabla_{\bm{\theta}}^2 \Loss(\bm{\theta})\big|_{\thetahat}$ gives the Gaussian approximation $\log\bm{\theta} \sim \N(\log\thetahat, \mathbf{H}^{-1})$.
\end{enumerate}
It should be noted that each callback redefines the loss function, so the optimization is performed on a changing objective. A visible consequence, is that the loss can abruptly jump after a callback, as seen in \cref{fig:single_loop}c. Nevertheless the distance of the estimates from the true parameter values keeps falling, indicating that the optimizer is making progress despite the non-monotonic loss.

\begin{figure}[!t]
    \centering
    \includesvg[inkscapearea=page]{figures/generated/fig_dataset_legend_col}\\[-1pt]
    \subfloat[]{\includesvg[inkscapearea=page]{figures/generated/fig_dataset_loadstep}\label{fig:dataset_loadstep}}\\[-2pt]
    \subfloat[]{\includesvg[inkscapearea=page]{figures/generated/fig_dataset_sinvref}\label{fig:dataset_sinvref}}\\[2pt]
    \subfloat[]{\footnotesize
      \sisetup{table-format=3.1, round-mode=places, round-precision=1}%
      \begin{tabular}{@{} l S l | l S l @{}}
      \toprule
      Parameter & {Value} & Unit & Parameter & {Value} & Unit \\
      \midrule
      $V_{\mathrm{in}}$     & 48.0  & \unit{V}            & $R_C$                 & 201.0 & \unit{\milli\ohm} \\
      $V_{\mathrm{ref}}$    & 24.0  & \unit{V}            & $R_{\mathrm{ds,on}}$ & 221.0 & \unit{\milli\ohm} \\
      $f_{\mathrm{sw}}$     & 20.0  & \unit{kHz}          & $V_F$                 & 1.0   & \unit{V} \\
      \addlinespace
      $L$                   & 725.0 & \unit{\micro\henry} & $R_1$                 & 10.2  & \unit{\ohm} \\
      $R_L$                 & 314.0 & \unit{\milli\ohm}   & $R_2$                 & 3.1   & \unit{\ohm} \\
      $C$                   & 164.5 & \unit{\micro\farad} & $R_3$                 & 6.1   & \unit{\ohm} \\
                            &       &                     & $R$                   & 5.0   & \unit{\ohm} \\
      \bottomrule
      \end{tabular}\label{fig:sim_specs}}
    \caption{Simulated benchmarks: (a)~load steps and (b)~sinusoidal reference, noiseless
    records (left) and five switching periods at 10~\glspl{LSB} with the samples of each
    strategy (right); (c)~simulation parameters, with the loads $R_1$--$R_3$ of (a) and $R$
    of (b).}
    \label{fig:sim_benchmarks}
\end{figure}
\section{Validation on Simulated Data}
\label{sec:val_simulated}
This section validates the estimator on simulated data with a known ground truth. The simulated converter has the parameters listed in the table in \cref{fig:sim_specs}; $i_L$ and $v_o$ are corrupted by additive white Gaussian noise. Its standard deviation is expressed in units of the \gls{LSB} of a 12-bit \gls{ADC}, a resolution common in the microcontrollers used for converter control~\cite{ti_tms320_datasheet}, with current and voltage ranges of \qty{10}{A} and \qty{30}{V}; with these settings, the \glspl{LSB}\ correspond to roughly \qty{2.4}{mA} and \qty{7.3}{mV}, respectively. Three levels are used, 1, 5 and 10~\glspl{LSB}, the same as in~\cite{Shuai2022}, which allows a direct comparison with that work (\cref{subsec:sim_sota}). They range from an acquisition limited by the resolution of the \gls{ADC} to a harsh case. Relative to the switching ripple, about \qty{0.24}{A} and \qty{47}{mV} RMS on both benchmarks, these levels correspond to \glspl{SNR}~\eqref{eq:snr} of approximately 40, 26 and \qty{20}{dB} on the current and 16, 2 and \qty{-4}{dB} on the voltage. \cref{subsec:transients} presents the two transient types used as benchmarks, \cref{subsec:sim_single_run} shows what the estimator delivers from one record, \cref{subsec:sim_mc} tests over repeated noise realizations whether its estimates are unbiased and its uncertainty correct, and what ignoring the residual correlation costs, and \cref{subsec:sim_sota} compares the estimator with the state of the art.

\subsection{Transient Types}
\label{subsec:transients}
\begin{figure*}[!tp]
    \centering
    \subfloat[]{\includesvg[inkscapearea=page]{figures/generated/fig_single_run_marginals}\label{fig:single_run_marginals}}\hfill
    \subfloat[]{\includesvg[inkscapearea=page]{figures/generated/fig_single_run_correlation}\label{fig:single_run_correlation}}\\[2pt]
    \subfloat[]{\footnotesize\setlength{\tabcolsep}{3.2pt}%
      \begin{tabular}{@{}l r r c r r c r r c r@{}}
  \toprule
  Parameter & True & \multicolumn{3}{c}{1 LSB} & \multicolumn{3}{c}{5 LSB} & \multicolumn{3}{c}{10 LSB} \\
  \cmidrule(lr){3-5} \cmidrule(lr){6-8} \cmidrule(lr){9-11}
   & & Est. & 95.4\,\% CI ($2\sigma$) & CV [\%] & Est. & 95.4\,\% CI ($2\sigma$) & CV [\%] & Est. & 95.4\,\% CI ($2\sigma$) & CV [\%] \\
  \midrule
  $L$ [\unit{\micro\henry}] & 725 & 724.81 & $[724.21,\,725.41]$ & 0.042 & 727.6 & $[724.6,\,730.6]$ & 0.20 & 726.9 & $[721.1,\,732.8]$ & 0.40 \\
  $R_L$ [\unit{\milli\ohm}] & 314 & 313.0 & $[308.0,\,318.1]$ & 0.81 & 301 & $[277,\,327]$ & 4.1 & 347 & $[301,\,400]$ & 7.1 \\
  $C$ [\unit{\micro\farad}] & 164.5 & 164.56 & $[164.44,\,164.68]$ & 0.036 & 164.05 & $[163.48,\,164.62]$ & 0.17 & 164.3 & $[163.1,\,165.4]$ & 0.34 \\
  $R_C$ [\unit{\milli\ohm}] & 201 & 201.60 & $[200.68,\,202.53]$ & 0.23 & 200.4 & $[195.9,\,205.0]$ & 1.1 & 197.8 & $[189.0,\,207.0]$ & 2.3 \\
  $R_{\mathrm{ds,on}}$ [\unit{\milli\ohm}] & 221 & 224.0 & $[215.3,\,233.0]$ & 2.0 & 247 & $[208,\,295]$ & 8.8 & 176 & $[108,\,287]$ & 25 \\
  $V_F$ [\unit{\volt}] & 1 & 1.048 & $[0.962,\,1.141]$ & 4.3 & 1.18 & $[0.82,\,1.72]$ & 19 & 1.26 & $[0.63,\,2.54]$ & 36 \\
  $V_{\mathrm{in}}$ [\unit{\volt}] & 48 & 48.044 & $[47.960,\,48.129]$ & 0.088 & 48.17 & $[47.75,\,48.59]$ & 0.43 & 48.28 & $[47.45,\,49.13]$ & 0.87 \\
  $R_1$ [\unit{\ohm}] & 10.2 & 10.20049 & $[10.19966,\,10.20132]$ & 0.0041 & 10.1995 & $[10.1955,\,10.2036]$ & 0.020 & 10.2029 & $[10.1946,\,10.2112]$ & 0.041 \\
  $R_2$ [\unit{\ohm}] & 3.1 & 3.099973 & $[3.099874,\,3.100073]$ & 0.0016 & 3.10038 & $[3.09989,\,3.10087]$ & 0.0079 & 3.10079 & $[3.09981,\,3.10176]$ & 0.016 \\
  $R_3$ [\unit{\ohm}] & 6.1 & 6.10000 & $[6.09969,\,6.10031]$ & 0.0026 & 6.1007 & $[6.0991,\,6.1022]$ & 0.013 & 6.0992 & $[6.0961,\,6.1023]$ & 0.025 \\
  \bottomrule
\end{tabular}
\label{fig:single_run_table}}
    \caption{Estimation from one record: the load-step benchmark sampled once per dwell, with one noise
    realization at each of 1, 5 and 10~\glspl{LSB}: (a)~posterior marginals, (b)~correlations,
    and (c)~\gls{MAP}, $\pm 2\sigma$ (95.4\,\%) credible interval and coefficient of variation
    of every parameter.}
    \label{fig:single_run}
\end{figure*}
The estimator needs a record in which the converter is excited, so that its parameters leave a trace in the waveforms. Two kinds of transient are used as benchmarks: load steps and sinusoidal reference (\cref{fig:sim_benchmarks}).

\paragraph*{Load steps} The three load-step records of~\cite{Shuai2022} switch the load between \qty{10.2}{\ohm}, \qty{6.1}{\ohm} and \qty{3.1}{\ohm} (\cref{fig:dataset_loadstep}), with the reference held constant. Load steps are widely used in the literature, which makes them a useful benchmark, and they give the easier estimation problem, since each step moves the converter far from its previous operating point. This results in estimated posterior distributions that are narrower than under the sinusoidal reference (\cref{subsec:sim_mc}). Their drawback appears on real hardware, where the effective parameters of real components change with the operating condition and a record that spans several operating points would yield some average over those points.

\paragraph*{Sinusoidal reference} One record at $R = \qty{5}{\ohm}$ modulates the reference instead, $V_\mathrm{ref}(t) = 24 + 2.4\sin(2\pi \cdot \qty{250}{Hz} \cdot t)$~V, which excites the converter without switching the load (\cref{fig:dataset_sinvref}). The modulation interferes with the control loop, but only briefly, for the few tens of milliseconds of the record, and with a slow, small modulation, which is acceptable in most applications. It gives the harder estimation problem, since with a single load the duty cycle varies little, and some parameters are less constrained (\cref{subsec:sim_mc}). In exchange, a small and slow modulation keeps the converter within a narrow band of operating points, so that the effective parameters of a real board stay around one value; this is why the hardware study of \cref{sec:val_hardware} uses it.

\subsection{Estimation from One Record}
\label{subsec:sim_single_run}

We now apply the estimator to a synthetic dataset and analyze the output. \Cref{fig:single_run} shows what the estimator delivers, for the load-step benchmark sampled once per dwell, with one noise realization at each of 1, 5 and 10~\glspl{LSB}.

\subsubsection{Estimates and Their Uncertainty}
\label{subsubsec:single_run_estimates}
\Cref{fig:single_run_marginals} shows the Laplace posterior of every parameter at each noise level: each density is scaled to unit peak, and below it are the \gls{MAP} estimate and its $\pm 2\sigma$ interval in log space at each noise level, with the true value dotted. The table in \cref{fig:single_run_table} lists the estimates, their $\pm 2\sigma$ (95.4\,\%) credible intervals, the intervals drawn in \cref{fig:single_run_marginals}, and their coefficients of variation.

The coefficient of variation (CV) is the posterior standard deviation of a parameter $\theta$ relative to its posterior mean,
\begin{equation}
    \mathrm{CV} = \frac{\sqrt{\mathrm{Var}[\theta]}}{\mathbb{E}[\theta]}.
    \label{eq:cv}
\end{equation}
Being dimensionless, it places parameters of different units and magnitudes on one scale.

The posteriors narrow as the noise falls, and all 30 intervals contain the true value; the farthest, $L$ at 5~\glspl{LSB}, lies 1.8 standard deviations away. The CV differs by orders of magnitude between parameters. At 10~\glspl{LSB}, it is below \qty{0.05}{\percent} for the load resistances and below \qty{1}{\percent} for $L$, $C$ and $V_\mathrm{in}$; it is \qty{2.3}{\percent} for $R_C$ and \qty{7}{\percent} for $R_L$, and reaches \qty{25}{\percent} for $R_\mathrm{ds,on}$ and \qty{36}{\percent} for $V_F$.

\subsubsection{Resolution of the Estimator}
\label{subsubsec:resolution}
For condition monitoring, the smallest change of a component that the estimator can tell apart from the noise is an important metric, which corresponds to the \emph{resolution} of the estimator. A change is detected by comparing two estimates of the same parameter, from records taken at different times, each with its own posterior. If the two estimates are independent and, for simplicity, have the same CV, each of them carries a relative uncertainty of one CV. Their uncertainties add in variance, $\mathrm{CV}^2 + \mathrm{CV}^2$, so the relative uncertainty of their difference is, to first order, $\sqrt{2}\,\mathrm{CV}$. Even if the component has not changed, the difference between the two estimates is therefore not zero, but fluctuates with a standard deviation of $\sqrt{2}\,\mathrm{CV}$. A measured change is credible only if it is large compared with this fluctuation, and it is called significant at level $\alpha$ when it exceeds
\begin{equation}
    \Delta_\mathrm{res}(\alpha) = z_{1-\alpha/2}\,\sqrt{2}\,\mathrm{CV},
    \label{eq:resolution}
\end{equation}
where $\alpha$ is the accepted probability of a false alarm, i.e.\ of declaring a change that noise alone has produced, and $z_{1-\alpha/2}$ is the number of standard deviations that sets it, the $(1-\alpha/2)$ quantile of the standard normal distribution: $z = 1.96$ for $\alpha = \qty{5}{\percent}$ and $z = 2.58$ for $\alpha = \qty{1}{\percent}$. The estimator resolves changes larger than $\Delta_\mathrm{res}$ at significance $\alpha$ and cannot resolve smaller ones. At the level of the $\pm 2\sigma$ intervals used throughout ($z = 2$, $\alpha = \qty{4.6}{\percent}$), $\Delta_\mathrm{res} = 2\sqrt{2}\,\mathrm{CV} \approx 2.8\,\mathrm{CV}$: at 10~\glspl{LSB}, for example, the CVs in the table in \cref{fig:single_run_table} (\qty{0.34}{\percent} for $C$ and \qty{36}{\percent} for $V_F$) give a resolution of about \qty{1}{\percent} for $C$, but about \qty{100}{\percent} for $V_F$, whose changes cannot be tracked from records of this quality. Since the CV comes from the posterior of a single record, the resolution is known before any change has occurred.

\subsubsection{Parameter Correlations}
\label{subsubsec:single_run_correlations}
\Cref{fig:single_run_correlation} shows the posterior correlations at 5~\glspl{LSB} (lower triangle, in percent), with the CV of each parameter on the diagonal. They show which parameters the record constrains mainly in combination. For example, $V_\mathrm{in}$ drives the inductor during the ON dwells and $V_F$ during the OFF dwells, so raising both together leaves the mean inductor voltage, and hence the record, nearly unchanged. Accordingly, the two are almost perfectly correlated ($+100\,\%$), and both are also correlated with $L$ ($+70\,\%$) and $R_C$ ($+52\,\%$). Similarly, $R_\mathrm{ds,on}$ and $R_L$ both drop voltage during the ON dwells, so the data fix mainly their sum ($-95\,\%$). The marginal credible intervals therefore give only part of the information. A simple consequence could be that in condition monitoring, a change in $V_F$, the less well determined of the two, is only meaningful when considered jointly with $V_\mathrm{in}$, which the joint posterior allows.

\subsection{Validation over Repeated Noise Realizations}
\label{subsec:sim_mc}

\begin{figure*}[!tp]
    \centering
    \includesvg[inkscapearea=page]{figures/generated/fig_mc_comparison}
    \caption{Uncertainty of the estimates in the six configurations: the load-step (LD) and sinusoidal-reference
    (SR) benchmarks with switching-point sampling (SP), one sample per dwell (PD) and one sample per period (PP), at each noise level. Density: the distribution of the
    estimate predicted by the Laplace posterior of one fit (median width over the 15 fits),
    about the true value; line: its $\pm 2\sigma$ credible interval; black ticks and white square: mean $\pm 2$ standard
    deviations and mean of the 15 estimates (square omitted where it does not fit between the ticks); the single load $R$ of the sinusoidal benchmark shares the
    panel of $R_1$.}
    \label{fig:mc_comparison}
\end{figure*}
\begin{figure*}[!tp]
    \centering
    \includesvg[inkscapearea=page]{figures/generated/fig_mc_calibration}
    \caption{Standardized errors $z$ of every run with (a)~the whitened and (b)~the unwhitened
    likelihood, pooled over the noise levels, per parameter and configuration, with their
    spread $\pm s_z$ and mean (white square); right: theory against each configuration.
    Parameters with a non-positive Laplace variance (no valid interval) are not drawn.}
    \label{fig:mc_calibration}
\end{figure*}

The quality of an estimator is set by its bias, the average offset of its estimates from the true parameters, and its variance, their scatter. Both are properties of the estimator over repeated experiments: a single record gives one estimate and cannot reveal them, whereas simulation can repeat the experiment with fresh noise against known true parameters. We therefore repeat every configuration, both transient types of \cref{fig:sim_benchmarks} under the three sampling strategies of \cref{subsec:sampling}, switching-point sampling (SP), one sample per dwell (PD) and one sample per period (PP), with 15 independent noise realizations at each of 1, 5 and 10~\glspl{LSB} (135 fits per benchmark, no prior). \Cref{fig:mc_comparison} compares the uncertainty of the estimates across the six configurations, and \cref{fig:mc_calibration} tests whether the estimates are unbiased and their stated uncertainty correct, through the standardized error $z = (\ln\hat\theta - \ln\theta^\star)/\hat\sigma_\mathrm{Lap}$ of each run, the error in units of the stated standard deviation, of which each pair of parameter and configuration has 45 values.

\subsubsection{Excitation and Sampling}
\label{subsubsec:mc_comparison}
\Cref{fig:mc_comparison} shows, for every parameter, noise level and configuration, the distribution of the estimate that the Laplace posterior of one fit predicts, centered on the true value, with a line of the same color below it for the predicted $\pm 2\sigma$ credible interval. The 15 estimates actually obtained are summarized by a white square at their mean and two black ticks at $\pm 2$ standard deviations from it. In every configuration the ticks fall close to the ends of the line, i.e.\ the predicted interval matches the observed scatter. The widths of the posteriors can therefore be compared directly across the six configurations.

The load steps determine most parameters better than the sinusoidal reference. With one sample per dwell, the sinusoidal reference widens the posteriors of $R_L$ and $R_\mathrm{ds,on}$ about five times, and those of $L$ and $R_C$ 2.5 to 3.7 times. Moving the samples from the switching instants to the dwell centers also has a cost. On the load steps, one sample per dwell widens the posteriors of $V_F$ and $V_\mathrm{in}$ by a factor of 4.4 to 4.7, and those of $R_L$, $R_C$ and $R_\mathrm{ds,on}$ by 1.7 to 2. Halving the samples to one per period widens every posterior by a nearly constant factor of 1.31 to 1.49 (median 1.40), close to the $\sqrt{2}$ expected when the number of samples is halved. The information is thus spread evenly over the samples. No parameter depends on sampling both dwells of a period, and the wider posteriors only reflect the smaller number of samples.

\subsubsection{Bias}
\label{subsubsec:unbiasedness}
The mean of the estimates is drawn as a white square in both \cref{fig:mc_comparison} and \cref{fig:mc_calibration}a; in the latter, it stays visibly close to zero in the standardized-error space. This suggests that the estimator is unbiased, which a $t$-test of zero mean for each parameter and configuration supports: corrected for the number of parameters tested (Holm's method~\cite{Holm1979}), it detects no bias at 95\,\% confidence in any of the six configurations.

\subsubsection{Uncertainty Quantification}
\label{subsubsec:uq_prediction}
For a correctly specified likelihood and informative data, the Bernstein--von Mises theorem~\cite[Ch.~10]{vanderVaart1998} states that the posterior of one record and the distribution of the estimate over repeated records are Gaussians with the same covariance, the Laplace covariance of \cref{subsec:posterior_methods}:
\begin{equation}
    p(\bm{\theta} \mid \D) \approx \N\bigl(\thetahat,\, \mathbf{H}^{-1}\bigr)
    \quad\text{and}\quad
    \thetahat \sim \N\bigl(\bm{\theta}^\star,\, \mathbf{H}^{-1}\bigr).
    \label{eq:bvm}
\end{equation}
The Laplace covariance $\mathbf{H}^{-1}$ of a single fit therefore predicts the scatter of the estimates over repeated records, which is what \cref{fig:mc_comparison} compares. The spread $s_z$ of each parameter's 45 standardized errors (bars in \cref{fig:mc_calibration}a) quantifies the agreement: the closer it is to one, the better the prediction, and it stays between 0.74 and 1.23 for every parameter, with 66--74\,\% of the standardized errors within $\pm 1$ and 95--98\,\% within $\pm 1.96$ in every configuration, against 68\,\% and 95\,\% for a calibrated posterior. A $\chi^2$ test of unit variance for each parameter and configuration, corrected as for the bias, confirms this: it detects no mismatch at 95\,\% confidence for any parameter in any of the six configurations.

The whitening is what makes this agreement possible. Dropping the lag-one covariance~\eqref{eq:gamma1_method}, i.e.\ treating the residuals as independent, misspecifies the likelihood, and the same tests then fail (\cref{fig:mc_calibration}b). The spread test flags 46 of the 52 parameters it can test, and the median $s_z$ drops from 0.93 to 0.25, so that most stated intervals are several times too wide and some too narrow. In addition, 41 of the 270 unwhitened fits end with a covariance that is not positive definite and state no interval at all, against none of the whitened fits.

\subsection{Comparison with Existing Benchmarks}
\label{subsec:sim_sota}
\begin{figure}[t]
    \centering
    \includesvg[inkscapearea=page]{figures/generated/fig_sota_comparison}\\
    \caption{The \gls{PINN} of~\cite{Shuai2022}, \gls{LS} and \gls{MLE} on the load-step
    benchmark with switching-point sampling: (a)~median relative error over the parameters;
    (b)~relative error of each parameter, in percent; (c)~wall time of one estimate.}
    \label{fig:accuracy_method_comparison}
\end{figure}

\subsubsection{Theoretical Accuracy Limit}
\label{subsubsec:efficiency}
The tests of \cref{subsec:sim_mc} support that the estimate is the \gls{MLE}, or the \gls{MAP} estimate when a prior is used. Its loss is the likelihood of all samples of both channels, it shows no bias, and its Laplace covariance, the inverse curvature of that likelihood, predicts the scatter of the estimates; the scatter is therefore at the Cram\'er--Rao bound (\cref{subsec:loss_meaning}). This has a strong consequence: \emph{with the same data, and without additional knowledge such as an informative prior, no unbiased estimator can outperform the \gls{MLE} on average.} It should be noted that the statement holds asymptotically, and not along the flat directions of the loss function. The comparison below puts it to the test against existing methods.

\subsubsection{Comparison with Existing Estimators}
\label{subsubsec:sota_setup}
We compare three estimators on the load-step benchmark with switching-point sampling. This setting is chosen since it matches the setting of~\cite{Shuai2022}:
\begin{itemize}
    \item the \gls{PINN}-based method of~\cite{Shuai2022}, taken as a well-established baseline;
    \item \gls{LS}: the RK4 forward model of \cref{subsec:buck_model} and the optimizer of \cref{subsec:optimizer}, minimizing the sum of squared residuals $\tfrac{1}{2}\sum_n \|\mathbf{r}_n\|^2$, i.e.\ with the residuals of both channels independent. This is the most commonly used loss function for parameter estimation.
    \item \gls{MLE}: the same forward model and optimizer with the likelihood~\eqref{eq:loss_general}, whitened as in \cref{subsec:whitening}, i.e.\ the complete pipeline.
\end{itemize}
All three run on the same machine, one estimate at a time. At each noise level (1, 5 and 10~\glspl{LSB}), three records are generated with different noise realizations, and every estimator is applied to these same three records. \Cref{fig:accuracy_method_comparison} reports the median of the three runs. The \gls{PINN} and \gls{LS} minimize the same unweighted loss, so their comparison isolates the forward model and the optimizer; \gls{LS} and \gls{MLE} share both, so theirs isolates the likelihood.

At 1~\gls{LSB} the three are comparable, with median errors over the parameters of \qtyrange{0.03}{0.21}{\percent}. As the noise grows, only \gls{MLE} keeps its accuracy: at 10~\glspl{LSB} its median error is \qty{0.33}{\percent}, against \qty{1.9}{\percent} for the \gls{PINN} and \qty{3.0}{\percent} for \gls{LS}, and the gap is largest on the weakly identified parameters: $V_F$ is estimated to within \qty{8}{\percent} by \gls{MLE}, while the \gls{PINN} misses it by \qty{91}{\percent} and \gls{LS} drives it to zero. \gls{LS} matches the accuracy of the \gls{PINN} about ten times faster, in \qty{1.8}{min} against 17 to \qty{19}{min}; the whitening adds about a minute and a half, \qty{3.4}{min} in all, and is what brings the accuracy at high noise. The \gls{MLE} has the smallest median error at every noise level in line with the theoretical accuracy prediction, and the few cells of \cref{fig:accuracy_method_comparison}b where a baseline comes closer, such as $R_2$ at 1~\gls{LSB}, are the chance wins on individual records that the statement allows.

\section{Validation on Hardware Data}
\label{sec:val_hardware}
This section tests the estimator on a laboratory buck converter, with two of the sampling strategies of \cref{subsec:sampling}, namely one sample per dwell and one sample per switching period.
The estimates are compared with a reference value or range for each parameter. Where possible, the reference is measured directly; otherwise, a reference range is obtained from independent estimates which are possible only from the densely sampled ripple shape, at \qty{6.25}{MS/s}, a rate that is not available in practical operation. We show that the intervals returned by the estimator agree with these references.

\Cref{subsec:hw_setup} describes the bench, the dataset and the datasheet prior, \cref{subsec:hw_misspecification}
where the board departs from the converter model, \cref{subsec:hw_inductor}
the iron-powder-core inductor model that is used, and \cref{sec:component_params} the reference values.
\Cref{subsec:hw_estimation,subsec:hw_consistency} compare the estimates with these references on one record and over ten records. Finally, \cref{subsec:hw_tracking} shows how the estimator can be used to track a change of the output capacitor, and \cref{subsec:hw_limitations} states the limitations.

\subsection{Setup, Dataset and Datasheet Prior}
\label{subsec:hw_setup}
\begin{figure*}[!t]
    \centering
    \subfloat[]{{\input{figures/generated/fig_lab_setup_photo.tex}}\label{fig:lab_setup_photo}}\hfill
    \subfloat[]{{\definecolor{ink}{HTML}{1F1F1F}\definecolor{mute}{HTML}{6E6E6E}%
\definecolor{cp}{HTML}{0072B2}\definecolor{cpd}{HTML}{00507F}%
\def\lsbd{\fontsize{7}{8.2}\selectfont}%
\def\lssm{\fontsize{6.3}{7.2}\selectfont}%
\begin{tikzpicture}[line width=0.45pt, color=ink, >={Stealth[length=3pt,width=2.6pt]},
    every node/.style={font=\lsbd, text=ink},
    box/.style={draw=ink!55, fill=ink!5, rounded corners=1.5pt, align=center, inner sep=1.5pt,
                minimum height=0.56cm},
    chain/.style={box, minimum width=1.45cm, text width=1.4cm},
    pr/.style={box, draw=cp!70, fill=cp!7, minimum width=2.05cm, text width=2.0cm},
    sig/.style={text=cp},
    dly/.style={font=\lssm, text=mute, above=-0.5pt, inner sep=1pt}]

\node[box, text width=1.7cm] (sup) at (0.88,4.35) {DC supply, 12\,V};
\node[box, text width=1.7cm] (mcu) at (0.88,3.35) {controller\\{\lssm TMS320F28379D}};
\node[box, minimum width=1.8cm, minimum height=1.6cm] (pwr) at (3.25,3.85)
  {buck power\\stage\\[1pt]$S$, $D$, $L$, $C$};
\node[box, minimum width=1.3cm] (load) at (5.3,4.35) {load bank\\$R$};
\draw[->] (sup) -- node[above, inner sep=1pt]{$V_\mathrm{in}$} (sup -| pwr.west);
\draw[->] (mcu) -- node[above, inner sep=1pt, font=\lssm]{PWM} (mcu -| pwr.west);
\draw[->] (pwr.east |- load) -- (load);

\def\ya{2.3}\def\yb{1.52}\def\yc{0.74}\def\yd{-0.04}
\def\xa{5.2}\def\xb{6.85}\def\xs{8.8}
\coordinate (ti) at (3.9,3.05); \coordinate (tv) at (3.25,3.05); \coordinate (tw) at (2.6,3.05);
\node[chain] (i1) at (\xa,\ya) {CT415 TMR\\sensor};
\node[chain] (i2) at (\xb,\ya) {OPA1652\\buffer};
\node[chain] (v1) at (\xa,\yb) {200k/10k\\divider};
\node[chain] (v2) at (\xb,\yb) {AMC1311B\\isolated $\Delta\Sigma$};
\node[chain] (w1) at (\xa,\yc) {3:1\\attenuator};
\node[chain, dashed] (f1) at (\xa,\yd) {clamp,\\OPA1652};
\node[chain, dashed] (f2) at (\xb,\yd) {ISO224B\\isolated, $\times\frac13$};
\draw[->] (ti) -- (ti |- 0,\ya) -- (i1);
\draw[->] (tv) -- (tv |- 0,\yb) -- (v1);
\draw[->] (tw) -- (tw |- 0,\yc) -- (w1);
\draw[->, dashed] (tw |- 0,\yc) -- (tw |- 0,\yd) -- (f1);
\fill (tw |- 0,\yc) circle (0.6pt);
\node[sig, anchor=south west, inner sep=1.5pt] at (ti |- 0,\ya) {$i_L$};
\node[sig, anchor=south west, inner sep=1.5pt] at (tv |- 0,\yb) {$v_o$};
\node[sig, anchor=south west, inner sep=1.5pt] at (tw |- 0,\yc) {$v_\mathrm{sw}$ {\lssm(diode)}};
\node[sig, anchor=south west, inner sep=1.5pt] at (tw |- 0,\yd) {{\lssm$V_F$ monitor}};
\draw[->] (i1) -- (i2); \draw[->] (v1) -- (v2); \draw[->, dashed] (f1) -- (f2);

\node[box, minimum width=0.95cm, minimum height=2.95cm, text width=0.9cm] (scope)
  at (\xs,{(\ya+\yd)/2}) {oscillo-\\scope\\[3pt]\lssm 8 bit};
\draw[->] (i2) -- (i2 -| scope.west);
\draw[->] (v2) -- (v2 -| scope.west);
\draw[->] (w1) -- (w1 -| scope.west);
\draw[->, dashed] (f2) -- (f2 -| scope.west);

\def\xp{10.65}
\node[pr] (ref) at (\xp,3.55) {dense record:\\reference values};
\node[pr] (p1) at (\xp,\ya) {Kaiser filter\\{\lssm zero phase, 123\,kHz}};
\node[pr] (p2) at (\xp,\yb) {switching instants\\from $v_\mathrm{sw}$};
\node[pr] (p3) at (\xp,\yc) {one sample per\\dwell or period};
\node[pr, draw=cpd, fill=cp!18, line width=0.6pt] (p4) at (\xp,\yd) {Bayesian\\estimator};
\draw[->] (scope.east |- p1) -- (p1);
\draw[->] (p1) -- (p2); \draw[->] (p2) -- (p3); \draw[->] (p3) -- (p4);
\draw[->] (p1) -- (ref);
\end{tikzpicture}%
}\label{fig:lab_signal_chain}}\\[2pt]
    \subfloat[]{\includesvg[inkscapearea=page]{figures/generated/fig_lab_dataset}\label{fig:lab_dataset}}
    \caption{The laboratory converter. (a)~Bench. (b)~Measurement chain from the circuit to the estimator. (c)~One record: the oscilloscope record (gray) and the samples the estimator receives, 159 with one sample per
    dwell (blue) or the 80 of them at the OFF-dwell centers with one sample per period (green rings).}
    \label{fig:lab_setup}
\end{figure*}
The test bench is the buck converter of \cref{fig:lab_setup}, controlled in closed loop by a Texas Instruments
TMS320F28379D microcontroller~\cite{ti_tms320_datasheet} at a switching frequency of \qty{2.5}{kHz}. As discussed in \cref{subsec:non_intrusive}, both strategies use samples that a digital controller takes with mean-value sampling. Here, the signals are nevertheless recorded densely, with an 8-bit oscilloscope (Tektronix TBS2104B) at the sense outputs of the board
(\cref{fig:lab_signal_chain}), and the samples of both schemes are extracted from this dense record (\cref{fig:lab_dataset}). The signals $i_L$ and $v_o$ are low-pass filtered with a zero-phase Kaiser filter (cutoff \qty{123}{kHz}), the switching instants are read off the switch-node voltage.

Every record uses the sinusoidal reference of \cref{subsec:transients}, in which $V_\mathrm{ref}$ is modulated with a
\qty{1}{V} sinusoid around \qty{6}{V}, at a measured frequency of \qty{125}{Hz}, with a \qty{3.3}{\ohm} load. A small, slow
modulation keeps the board within a narrow band of operating points, so that the parameter values remain stable. A record spans \qty{32}{ms}, some 80 switching periods, and gives 159 samples with one sample per dwell, or 80 with one sample per period (\cref{fig:lab_dataset}). Ten records are collected under the same conditions, to verify the consistency of the estimates and the accuracy of their predicted uncertainty.

For parameters on which we have prior knowledge from the setup or from the datasheet, a lognormal prior is assigned,
$\ln\theta_j \sim \N(\ln\mu_{0,j},\,\sigma_{\log,j}^2)$, which enters~\eqref{eq:map_loss} as
$\sum_j (\ln\theta_j - \ln\mu_{0,j})^2 / (2\sigma_{\log,j}^2)$ (\cref{tab:lab_reference}). The inductor datasheet states the zero-bias small-signal inductance, not the parameter $L$ of the model of \cref{subsec:hw_inductor}, and only a maximum DC resistance; also, the capacitor datasheet does not state the \gls{ESR} at the switching frequency. Therefore, no prior is assigned to these parameters.

\begin{table}[!t]
    \caption{Datasheet Prior and Independent Reference}
    \label{tab:lab_reference}
    \centering
    \footnotesize
    \setlength{\tabcolsep}{3.5pt}
    \begin{tabular}{@{}l c c l l@{}}
    \toprule
     & \multicolumn{2}{c}{datasheet prior} & \multicolumn{2}{l}{reference} \\
    \cmidrule(lr){2-3}\cmidrule(l){4-5}
     & $\mu_0$ & $\sigma_{\log}$ & range & source \\
    \midrule
    $\bar L$         & --                      & --                 & 1008--1146\,\unit{\micro\henry} & dwell slopes \\
    $R_L$            & --                      & --                 & 180--500\,\unit{\milli\ohm}      & DC resistance \\
    $C$              & \qty{150}{\micro\farad} & \qty{20}{\percent} & 104.8--127.6\,\unit{\micro\farad} & frequency lines \\
    $R_C$            & --                      & --                 & 282--323\,\unit{\milli\ohm}      & frequency lines \\
    $R_\mathrm{ds,on}$ & \qty{45}{\milli\ohm}  & \qty{60}{\percent} & 45--65\,\unit{\milli\ohm}        & datasheet~\cite{mosfet_datasheet} \\
    $V_F$            & \qty{1.0}{\volt}        & \qty{20}{\percent} & 0.951--1.004\,\unit{\volt}       & monitor \\
    $V_\mathrm{in}$  & \qty{12}{\volt}         & \qty{5}{\percent}  & \qty{12}{\volt}                   & supply setting \\
    $R$              & \qty{3.3}{\ohm}         & \qty{20}{\percent} & 3.2--3.4\,\unit{\ohm}            & bench, $\pm$\qty{0.1}{\ohm} \\
    \bottomrule
    \end{tabular}
\end{table}

\subsection{System Model Misspecification}
\label{subsec:hw_misspecification}
The model of \cref{subsec:buck_model} idealizes the board, describing every component by one constant parameter.
Real components do not behave this way, and where the model cannot reproduce the behavior of the system that produced the data, it is misspecified~\cite{kennedy2001bayesian}. A misspecified model can still be fitted, and indeed some degree of misspecification is inevitable in any practical system, but the quality of the fit will worsen with respect to the idealized model. 

On this board, three departures are worth noting.
\begin{itemize}
    \item \emph{Inductor.} The cores are of iron powder, and the effective inductance changes considerably within every dwell. This behavior is a significant departure from the idealized constant inductance, so the model is extended with an inductance law (\cref{subsec:hw_inductor}).
    \item \emph{Output capacitor.} It is an aluminum electrolytic, whose capacitance and series resistance depend on frequency. It therefore behaves differently for slow changes of the voltage and for the faster switching ripple.
    We accept a simplified model with one constant pair, $C$ and $R_C$, and assume that it captures the average behavior over the relevant frequency range.
    \item \emph{Diode.} Its forward drop rises with the current, but we again accept a simplified model and keep $V_F$ constant.
\end{itemize}

\subsection{Inductor Model}
\label{subsec:hw_inductor}
\begin{figure*}[!t]
    \centering
    \includesvg[inkscapearea=page]{figures/generated/fig_lab_inductor_hysteresis}
    \caption{From the hysteresis of the core to the inductance law. (a)~Major loop, magnetization curve and the
    minor loop the switching ripple traces at the operating point (schematic). (b)~The ripple loop
    of~\eqref{eq:inductance_law} with the held shape, on one record; the aftereffect acts in time and
    is not part of the loop. (c)~The loop of (b) and the same loop for a smaller ($-$) and a larger ($+$) base
    inductance $L$, the only parameter of~\eqref{eq:inductance_law} that is estimated.}
    \label{fig:lab_inductor_hysteresis}
\end{figure*}
The filter inductance is three Signal Transformer HCTI-270-5.5 units in series~\cite{inductor_datasheet}, toroids
on iron-powder cores. \Cref{fig:lab_inductor_hysteresis} shows that as the current ripples around its mean value, the core traces a small (minor) loop inside its
hysteresis loop (\cref{fig:lab_inductor_hysteresis}a). The inductance the circuit sees is the slope of this small loop, which is the change of the flux linkage $\lambda$ per change of the current $i$,
$L_\mathrm{eff} = \mathrm{d}\lambda/\mathrm{d}i$, so that the inductor voltage is
$v_L = L_\mathrm{eff}\,\mathrm{d}i/\mathrm{d}t$. 

This slope is not constant, and three effects describe how it varies. The first two are properties of the loop itself, that is, of the path that $\lambda$ follows as the current ripples, whereas the third depends on time and cannot be read from the shape of the loop.
\begin{itemize}
\item \textit{Saturation}: The small loop becomes flatter at a higher current (\cref{fig:lab_inductor_hysteresis}a), reducing the effective inductance~\cite{imaoka2019powder}.
\item \textit{Rayleigh}: The loop is flattest right after a switching edge, where the current reverses, and becomes steeper the further the current moves from the reversal (\cref{fig:lab_inductor_hysteresis}b)~\cite{rayleigh1887}.
\item \textit{Aftereffect}: The response of the core is delayed for some tens of microseconds after each edge, temporarily lowering the inductance~\cite{preisach2017aftereffect}.
\end{itemize}

The inductance law approximates each effect with one term:
\begin{equation}
    L_\mathrm{eff}(i, t_r) = \frac{L + \beta\,|i - i^\star| - \gamma\,(i - i_\mathrm{ref})}{1 + \eta\, e^{-t_r/\tau}},
    \label{eq:inductance_law}
\end{equation}
where four coefficients set the shape of the law, namely $\gamma$ and $\beta$ for the saturation and the Rayleigh term, and the amplitude $\eta$ and the time constant $\tau$ of the aftereffect, with $\eta = \eta_\mathrm{on}$ after a turn-on edge and $\eta_\mathrm{off}$ after a turn-off edge. Also, we have the base inductance $L$, the mean current over the record $i_\mathrm{ref}$, the current at the last switching event $i^\star$ and the time since the switch $t_r$. With $\eta = \beta = \gamma = 0$ the law reduces to the constant inductance of \cref{subsec:buck_model}.

These shape coefficients are properties of the core material and must be characterized before the law can be used for estimation. We mimic this step by fitting them once to the dense record and using these values for all the fits. In particular, we obtain $\eta_\mathrm{on} = 0.33$, $\eta_\mathrm{off} = 0.17$, $\beta = \qty{300}{\micro\henry\per\ampere}$, $\tau = \qty{30}{\micro\second}$ and $\gamma = \qty{150}{\micro\henry\per\ampere}$.

Only $L$, which depends on the operating point, is estimated together with the other parameters. With the shape held, it tilts the loop as a whole and leaves its opening unchanged (\cref{fig:lab_inductor_hysteresis}c), so the estimation searches for the tilt that matches the record. For simplicity, we report the inductance averaged over the dwells of the record, $\bar L$, which has a clear physical meaning.

\subsection{Reference Values from the Dense Record}
\label{sec:component_params}
\begin{figure*}[!t]
    \centering
    \subfloat[]{\includesvg[inkscapearea=page]{figures/generated/fig_lab_ref_inductor}\label{fig:lab_ref_inductor}}\hfill
    \subfloat[]{\includesvg[inkscapearea=page]{figures/generated/fig_lab_ref_capacitor}\label{fig:lab_ref_capacitor}}
    \caption{References read off the dense samples of one record, which the estimator does not see.
    (a)~Inductance $v_L/(\mathrm{d}i_L/\mathrm{d}t)$ of every ON and OFF dwell; inset: one least-squares line per
    dwell. (b)~Capacitor: the measured output voltage and its spectrum, with its five tones (1--5); for each tone,
    the output voltage and the capacitor current over two periods, scaled to unit amplitude, and the $C$ and $R_C$
    it gives; the extreme tones set the reference range (shaded).}
    \label{fig:lab_references}
\end{figure*}
Because of the simple topology, the dense record (\qty{6.25}{MS/s}, about 1000 samples per dwell) can be analyzed
directly for some components, as shown in \cref{fig:lab_references}.
\begin{itemize}
    \item \emph{Inductor} (\cref{fig:lab_references}a): within a dwell the current is nearly a straight line, so
    $L = v_L/(\mathrm{d}i_L/\mathrm{d}t)$, with $v_L$ the inductor voltage and $i_L$ the inductor current, is read
    from the least-squares slope of every dwell. Neglecting the
    resistive drops makes the ON estimate read high and the OFF estimate read low, so their medians over the record bound the dwell-averaged inductance, $\bar L \in [1008, 1146]\,\mu\text{H}$, well above the \qty{810}{\micro\henry} of the datasheet.
    \item \emph{Capacitor} (\cref{fig:lab_references}b): an electrolytic capacitor behaves differently at
    different frequencies, so its reference is read in the frequency domain, one tone at a time. The record
    contains five dominant tones, namely the modulation at \qty{125}{Hz}, and the switching frequency at \qty{2.5}{kHz} with its
    harmonics up to \qty{10}{kHz}. A least-squares fit of the inductor current and the output voltage as sums of
    sinusoids at these frequencies gives their phasors $I_L$ and $V_o$ at each tone. The capacitor current is $I_C = I_L - V_o/R$, and the impedance of the capacitor branch gives both parameters:
    \begin{equation}
        \begin{gathered}
            Z_C = \frac{V_o}{I_L - V_o/R} = R_C + \frac{1}{j\omega C},\\
            C = -\frac{1}{\omega \operatorname{Im} Z_C}, \qquad R_C = \operatorname{Re} Z_C,
        \end{gathered}
        \label{eq:capacitor_lines}
    \end{equation}
    where $Z_C$ is the complex impedance of the capacitor branch, $\omega$ the angular frequency of the tone. Each tone can thus be used to estimate its own $C$ and $R_C$, although at
    \qty{125}{Hz}, $R_C$ cannot be resolved because the reactance of the capacitor is about thirty times larger. The estimates for each tone are reported in \cref{fig:lab_ref_capacitor} and used to set a reference range $C \in [104.8, 127.6]\,\mu\text{F}$ and $R_C \in [282, 323]\,\mathrm{m}\Omega$.
    \item \emph{Diode}: a dedicated monitor of the switch node reads the forward drop
    of every OFF dwell, which varies with the current over the record; the model's constant $V_F$ is compared with this range, $V_F \in [0.951, 1.004]\,\text{V}$.
    \item \emph{Other parameters}: the supply and the load are set ($V_\mathrm{in} = \qty{12}{\volt}$, $R$ to $3.3\,\Omega$ from the bench multimeter, but we allow of $0.1\,\Omega$ uncertainty). $R_\mathrm{ds,on}$ cannot be measured in circuit, so its reference is the range around its datasheet values, \qtyrange{45}{65}{\milli\ohm}~\cite{mosfet_datasheet}. For $R_L$, only the datasheet DC resistance of \qty{180}{\milli\ohm} is available, which is a lower bound, since in operation $R_L$ also carries the core loss. The reference range of $R_L$ is \qtyrange{180}{500}{\milli\ohm}, where \qty{500}{\milli\ohm} is a reasonable upper bound.
\end{itemize}
All reference values are collected in \cref{tab:lab_reference}.

\subsection{Estimation Results on the Hardware Dataset}
\label{subsec:hw_estimation}
\begin{figure*}[!tp]
    \centering
    \subfloat[]{\includesvg[inkscapearea=page]{figures/generated/fig_lab_posteriors}\label{fig:lab_posteriors_densities}}\\[2pt]
    \subfloat[]{\footnotesize\setlength{\tabcolsep}{2pt}\renewcommand{\arraystretch}{0.95}%
      \definecolor{lsOrange}{HTML}{D55E00}%
\definecolor{lsGreen}{HTML}{009E73}%
\definecolor{lsPrior}{HTML}{F1F4F8}%
\colorlet{lsIn}{lsGreen!16}%
\colorlet{lsOut0}{lsOrange!10}%
\colorlet{lsOut1}{lsOrange!22}%
\colorlet{lsOut2}{lsOrange!38}%
\definecolor{lsBlue}{HTML}{000000}%
\colorlet{lsCv1}{lsBlue!10}%
\colorlet{lsCv2}{lsBlue!18}%
\colorlet{lsCv3}{lsBlue!30}%
\definecolor{lsDwellprior}{HTML}{0B4F8C}%
\definecolor{lsDwellnoprior}{HTML}{6FAEDB}%
\definecolor{lsPeriodprior}{HTML}{1B7A50}%
\definecolor{lsPeriodnoprior}{HTML}{86CFA8}%
\setlength{\aboverulesep}{0pt}\setlength{\belowrulesep}{0pt}%
\setlength{\tabcolsep}{1.6pt}%
\renewcommand{\arraystretch}{1.2}%
\begin{tabularx}{\textwidth}{@{} l l *{3}{>{\columncolor{lsPrior}\centering\arraybackslash}X}>{\columncolor{lsPrior}}c *{3}{>{\centering\arraybackslash}X}c *{3}{>{\columncolor{lsPrior}\centering\arraybackslash}X}>{\columncolor{lsPrior}}c *{3}{>{\centering\arraybackslash}X}c}
  \toprule
  & & \multicolumn{8}{>{\columncolor{lsDwellprior!12}}c}{\textbf{one sample per dwell}} & \multicolumn{8}{>{\columncolor{lsPeriodprior!12}}c}{\textbf{one sample per period}} \\
  \cmidrule(lr){3-10} \cmidrule(l){11-18}
  & & \multicolumn{4}{>{\columncolor{lsPrior}}c}{with the datasheet prior} & \multicolumn{4}{c}{without prior} & \multicolumn{4}{>{\columncolor{lsPrior}}c}{with the datasheet prior} & \multicolumn{4}{c}{without prior} \\
  \cmidrule(lr){3-6} \cmidrule(lr){7-10} \cmidrule(lr){11-14} \cmidrule(l){15-18}
  Parameter & Reference & \makecell[cb]{Est.} & \makecell[cb]{CV\\{}[\%]} & \makecell[cb]{Acc.\\{}[\%]} & \makecell[cb]{$2\sigma$\\$\cap$\,ref.} & \makecell[cb]{Est.} & \makecell[cb]{CV\\{}[\%]} & \makecell[cb]{Acc.\\{}[\%]} & \makecell[cb]{$2\sigma$\\$\cap$\,ref.} & \makecell[cb]{Est.} & \makecell[cb]{CV\\{}[\%]} & \makecell[cb]{Acc.\\{}[\%]} & \makecell[cb]{$2\sigma$\\$\cap$\,ref.} & \makecell[cb]{Est.} & \makecell[cb]{CV\\{}[\%]} & \makecell[cb]{Acc.\\{}[\%]} & \makecell[cb]{$2\sigma$\\$\cap$\,ref.} \\
  \midrule
  $\bar L$ [\unit{\micro\henry}] & 1008--1146 & 1062 & 0.87 & \cellcolor{lsIn}in & \cellcolor{lsIn}\ding{51} & 1069 & 0.91 & \cellcolor{lsIn}in & \cellcolor{lsIn}\ding{51} & 1062 & 1.7 & \cellcolor{lsIn}in & \cellcolor{lsIn}\ding{51} & 1060 & 1.5 & \cellcolor{lsIn}in & \cellcolor{lsIn}\ding{51} \\
  $R_L$ [\unit{\milli\ohm}] & 180--500 & 254 & \cellcolor{lsCv2}10 & \cellcolor{lsIn}in & \cellcolor{lsIn}\ding{51} & 265 & \cellcolor{lsCv2}23 & \cellcolor{lsIn}in & \cellcolor{lsIn}\ding{51} & 270 & \cellcolor{lsCv2}30 & \cellcolor{lsIn}in & \cellcolor{lsIn}\ding{51} & 253 & \cellcolor{lsCv3}52 & \cellcolor{lsIn}in & \cellcolor{lsIn}\ding{51} \\
  $C$ [\unit{\micro\farad}] & 104.8--127.6 & 128.7 & 1.1 & \cellcolor{lsOut0}$+$0.9 & \cellcolor{lsIn}\ding{51} & 127.8 & 1.1 & \cellcolor{lsOut0}$+$0.2 & \cellcolor{lsIn}\ding{51} & 130.0 & 1.7 & \cellcolor{lsOut0}$+$1.9 & \cellcolor{lsIn}\ding{51} & 130.4 & 1.6 & \cellcolor{lsOut0}$+$2.2 & \cellcolor{lsIn}\ding{51} \\
  $R_C$ [\unit{\milli\ohm}] & 282--323 & 265 & \cellcolor{lsCv1}3.5 & \cellcolor{lsOut0}$-$5.9 & \cellcolor{lsIn}\ding{51} & 270 & \cellcolor{lsCv1}3.4 & \cellcolor{lsOut0}$-$4.1 & \cellcolor{lsIn}\ding{51} & 184 & \cellcolor{lsCv3}96 & \cellcolor{lsOut2}$-$35 & \cellcolor{lsIn}\ding{51} & 166 & \cellcolor{lsCv3}110 & \cellcolor{lsOut2}$-$41 & \cellcolor{lsIn}\ding{51} \\
  $R_\mathrm{ds,on}$ [\unit{\milli\ohm}] & 45--65 & 33 & \cellcolor{lsCv3}64 & \cellcolor{lsOut1}$-$28 & \cellcolor{lsIn}\ding{51} & -- & -- & -- & -- & 36 & \cellcolor{lsCv3}65 & \cellcolor{lsOut1}$-$19 & \cellcolor{lsIn}\ding{51} & -- & -- & -- & -- \\
  $V_F$ [\unit{\volt}] & 0.951--1.004 & 1.01 & \cellcolor{lsCv1}2.4 & \cellcolor{lsOut0}$+$1.0 & \cellcolor{lsIn}\ding{51} & 1.04 & \cellcolor{lsCv1}8.9 & \cellcolor{lsOut0}$+$3.2 & \cellcolor{lsIn}\ding{51} & 1.03 & \cellcolor{lsCv1}4.7 & \cellcolor{lsOut0}$+$3.0 & \cellcolor{lsIn}\ding{51} & 1.03 & \cellcolor{lsCv2}16 & \cellcolor{lsOut0}$+$2.6 & \cellcolor{lsIn}\ding{51} \\
  $V_\mathrm{in}$ [\unit{\volt}] & 12 & 11.98 & 0.83 & \cellcolor{lsOut0}$-$0.2 & \cellcolor{lsIn}\ding{51} & 12.05 & 1.0 & \cellcolor{lsOut0}$+$0.4 & \cellcolor{lsIn}\ding{51} & 12.03 & \cellcolor{lsCv1}2.3 & \cellcolor{lsOut0}$+$0.3 & \cellcolor{lsIn}\ding{51} & 11.97 & \cellcolor{lsCv1}2.6 & \cellcolor{lsOut0}$-$0.3 & \cellcolor{lsIn}\ding{51} \\
  $R$ [\unit{\ohm}] & 3.20--3.40 & 3.352 & 0.50 & \cellcolor{lsIn}in & \cellcolor{lsIn}\ding{51} & 3.352 & 0.50 & \cellcolor{lsIn}in & \cellcolor{lsIn}\ding{51} & 3.336 & 0.90 & \cellcolor{lsIn}in & \cellcolor{lsIn}\ding{51} & 3.336 & 0.88 & \cellcolor{lsIn}in & \cellcolor{lsIn}\ding{51} \\
  \bottomrule
\end{tabularx}%
\label{fig:lab_summary_table}}
    \caption{Estimates from one record of the hardware dataset. (a)~Laplace posteriors with one sample per dwell
    (left) and one per period (right), with the datasheet prior (strong) and without prior (light), with their
    $\pm 2\sigma$ intervals and the shift introduced by the prior (arrows), against the references of
    \cref{tab:lab_reference} (orange). (b)~Estimate, CV, accuracy and reference overlap of the $\pm 2\sigma$
    interval for the four test cases. Since most references are ranges, the accuracy is ``in'' when the estimate lies
    within the range, and otherwise its signed relative distance from the nearest end of the range.}
    \label{fig:lab_posteriors}
\end{figure*}
\Cref{fig:lab_posteriors} shows the performance of the estimator on the hardware dataset in four test cases: the two sampling schemes, each with and without the datasheet prior of \cref{tab:lab_reference}. Each panel draws the posterior of one parameter, scaled to unit peak, with its median and $\pm 2\sigma$ interval, against its reference. The table in \cref{fig:lab_summary_table} lists the median, the CV and the accuracy of each estimate. Its last column checks whether the $\pm 2\sigma$ credible interval of the estimate overlaps the reference range. An overlap means that the estimate, with the uncertainty the estimator itself reports, is consistent with the reference, even when its median lies outside the range. A parameter whose $95\,\%$ interval spans more than a factor of 1000 is considered undetermined and marked with a dash, as is $R_\mathrm{ds,on}$ without the prior. Every other credible interval overlaps its reference range, under both sampling schemes and with or without the datasheet prior. The estimates are therefore consistent with the references within their reported uncertainty in all four test cases.

The shape of the posterior distribution shows how well each sampling scheme determines the parameters. The credible intervals with one sample per dwell are narrower than those with one sample per period, by a factor of 1.5 to 2 for $\bar L$, $C$ and $R$ and of about 3 for $V_\mathrm{in}$. The datasheet prior acts mostly on the parameters that the data alone constrain weakly. It narrows the intervals of $R_L$ and $V_F$ by factors of about 2 and 3.5 and determines $R_\mathrm{ds,on}$, without significantly altering the estimates of the well-constrained parameters $\bar L$, $C$ and $R$.

\subsection{Validation over Repeated Records}
\label{subsec:hw_consistency}
\begin{figure}[!t]
    \centering
    \includesvg[inkscapearea=page]{figures/generated/fig_lab_scatter_calibration}
    \caption{Standardized deviations $z$ of the ten records of the same excitation from their inverse-variance-weighted
    mean, with the datasheet prior, per parameter and sampling scheme, with their spread $\pm s_z$ and mean (white
    square); dotted: the range of $s_z$ expected by chance for ten records; below: theory against each scheme.}
    \label{fig:lab_scatter_calibration}
\end{figure}
Ten independent measurements were collected under the same conditions to test the stability and uncertainty quantification of the estimator on experimental data. Unlike in \cref{subsec:sim_mc}, the true values are now unknown, so each estimate is compared with the inverse-variance-weighted mean $\bar\theta$ of the ten records instead of the truth, and a bias cannot be tested. 

\subsubsection{Stability of the Estimates}
\label{subsubsec:hw_stability}
The estimator is run on each of the ten records with the same settings, and the ten estimates of each parameter
should agree with each other. As in \cref{subsec:hw_estimation}, two estimates are taken as consistent when their
$\pm 2\sigma$ credible intervals overlap. Ten records form 45 pairs, which for the eight parameters gives 360
pairs of estimates. With the datasheet prior and one sample per dwell, \qty{97.8}{\percent} of them overlap (352 of
360). Six parameters overlap in every pair, and the two exceptions are $V_F$, with \qty{96}{\percent} (43 of 45),
and $R_C$, with \qty{87}{\percent} (39 of 45). With one sample per period, whose intervals are wider, all 360 pairs
overlap.

\subsubsection{Uncertainty Quantification}
\label{subsubsec:hw_uq}
This test verifies how well the uncertainty stated by the posterior matches the actual variability of the
estimates. \Cref{fig:lab_scatter_calibration}, which is the hardware counterpart of
\cref{fig:mc_calibration}, shows the standardized deviations
$z_k = (\ln\hat\theta_k - \ln\bar\theta)/\hat\sigma_k$ of the ten records, with the weighted mean $\bar\theta$ in place of the unknown truth, and their spread $s_z$.

The table in \cref{fig:lab_scatter_calibration}, below the plot, gives the same statistics as for the simulated data and shows that the agreement is less uniform than in \cref{fig:mc_calibration}. The bars of most parameters are shorter than $\pm 1$, so their posteriors are slightly conservative, while the bar of $R_C$ is longer, so its posterior underestimates its variability. With one sample per period, the posteriors are more conservative, with \qty{89}{\percent} and \qty{99}{\percent} of the deviations within the two bounds and a spread of 0.73. A $\chi^2$ test of consistency makes this quantitative. It checks whether the ten estimates of a parameter scatter more than their stated uncertainties allow, through the sum $Q = \sum_k z_k^2$ over the records, which follows a $\chi^2$ distribution with nine degrees of freedom when they do not. At \qty{95}{\percent} confidence, corrected for the eight parameters by Holm's method as in \cref{subsubsec:unbiasedness}, it rejects only $R_C$ with one sample per dwell ($p = 0.007$), and no parameter with one sample per period. Overall, the posterior of a single record is a good, if slightly conservative, indicator of the variability of its estimates. Since the same test finds the posterior calibrated on simulated data, where the model is exact, the departures found here can be attributed to the mismatch between the board and its model (\cref{subsec:hw_misspecification}).

\subsection{Tracking a Change of the Output Capacitor}
\label{subsec:hw_tracking}
\begin{figure*}[!tp]
    \centering
    \includesvg[inkscapearea=page]{figures/generated/fig_lab_tracking_combined}
    \caption{Capacitor sweep and condition monitoring of the output capacitor, with the datasheet prior (strong) and
    without prior (light): (a)~estimates of every parameter as the extra capacitors are removed against the references (orange) (b)~estimate of $C$, (c)~slope of its weighted linear trend, (d)~significance of the trend in every parameter, with a red ring on the first inspection above the warning level, and, beside it, the posterior of its final slope with the datasheet prior with the region of slopes considered negligible in light gray.}
    \label{fig:lab_tracking}
\end{figure*}
Aging lowers the capacitance of aluminum electrolytics and raises their \gls{ESR}, which makes the output capacitor a
natural target for condition monitoring. To emulate a change, four removable \qty{3.3}{\micro\farad} capacitors
(EGT335M1HD11RR~\cite{small_c_datasheet}, \qty{2.72}{\micro\farad} each on the LCR bridge) are connected in parallel
with the output capacitor and removed one at a time, and each of the five configurations is estimated independently,
as an inspection $k = 0, \ldots, 4$. After inspection $k$, a straight line $\log\hat\theta_j = a + s\,j$ is fitted to the log estimates of the inspections $j = 0, \ldots, k$ by weighted least squares, each weighted by the inverse of its
posterior variance. Here $a$ is the intercept, the slope $s$ is the relative change per inspection, and $\sigma_s$ is the standard deviation of the slope given by the fit. The significance of the trend is
\begin{equation}
    c_k = 1 - p_k, \qquad p_k = 2\,\Phi\!\left(-|s|/\sigma_s\right),
    \label{eq:change_evidence}
\end{equation}
with $\Phi$ the standard normal distribution function and $p_k$ the false-alarm probability, and a warning is raised
when $c_k > \qty{99}{\percent}$. 
Since every inspection enters with its own posterior variance, the test inherits the
calibration of the posterior.

\Cref{fig:lab_tracking}a shows the estimates over the sweep against the references derived from the dense records. With one sample per dwell, $C$ falls by $2.0 \pm 0.3\,\%$ per inspection with the datasheet prior ($-2.68 \pm 0.44$~\unit{\micro\farad}, against the \qty{2.72}{\micro\farad} removed) and $2.1 \pm 0.3\,\%$ without (\cref{fig:lab_tracking}b,c). It raises a warning at inspection 2 without ($c_2 = \qty{99.2}{\percent}$), after two removals, \qty{5.4}{\micro\farad} or about \qty{4}{\percent} of the capacitance. $R_C$ rises by $5.1 \pm 1.1\,\%$ per inspection, the largest change in the figure, and warns at inspection 2 in both arms. One sample per period carries less information per inspection. There, $C$ falls by $1.5 \pm 0.5\,\%$ per inspection and warns at the last inspection in both arms ($c_4 =\qty{99.8}{\percent}$ with the prior), while $R_C$, barely determined, raises no warning. 

The performance of the monitoring method can be assessed based on the resolution of the parameter estimates presented in~\eqref{eq:resolution}. Indeed, over the sweep, the CV of $C$ is about \qty{1.0}{\percent} with one sample per dwell and \qty{1.5}{\percent} with one per period, so by~\eqref{eq:resolution} the \textit{estimator resolution} on $C$ is about \qty{2.9}{\percent} and \qty{4.2}{\percent}, respectively. One removal, \qty{2.1}{\percent}, is below both, and two removals, \qty{4.2}{\percent}, exceed the one-sample-per-dwell resolution but not the one-sample-per-period resolution. 

It should be noted that none of the six parameters that did not change raises a warning, in any test case or inspection (\cref{fig:lab_tracking}d). This confirms that the monitoring method is specific to the parameters that actually vary, and does not produce false alarms for the stable ones.

\subsection{Limitations}
\label{subsec:hw_limitations}
The results of this paper hold within the following limitations, which also set the next steps.
\begin{itemize}
    \item \emph{Shape of the inductance law.} The estimator cannot determine the shape parameters
    of~\eqref{eq:inductance_law}, $\eta_\mathrm{on}$, $\eta_\mathrm{off}$, $\beta$, $\tau$ and $\gamma$, from one
    sample per dwell or per period. Estimating them together with the circuit parameters leaves many parameters weakly
    determined, and the optimizer does not converge reliably. The shape is therefore determined beforehand from dense
    records and held (\cref{subsec:hw_inductor}), and the estimates are conditional on it.
    \item \emph{Convergence along flat directions.} $V_\mathrm{in}$, $R_L$, $R_\mathrm{ds,on}$ and $V_F$ all act as
    drops on the inductor voltage, so the record constrains mainly their combination and the loss is nearly flat
    along one direction. There the optimizer of \cref{subsec:optimizer} advances slowly, and some of the hardware
    fits end at their iteration budget before reaching the optimum. Along this direction, their estimates are the end
    points of the optimizer, and their intervals come from the curvature at these points. An optimizer suited to flat
    directions is needed to remove this dependence.
    \item \emph{Effect of model misspecification.} The departures from calibration on hardware are attributed to the
    mismatch between the board and its model (\cref{subsubsec:hw_uq}), but this attribution is not tested directly. A systematic study on simulated data, with a known model error of adjustable size, is needed to quantify its
    effect on the bias of the estimates and on the predicted uncertainty.
    \item \emph{Converter topology.} The estimator is tested only on the buck converter, a simple topology. Its performance on other topologies remains to be verified, such as three-phase inverter systems \cite{Kong2026}.
\end{itemize}

\section{Conclusion}
\label{sec:conclusion}
This paper presents a method to non-intrusively estimate the circuit parameters of a power converter together with a calibrated posterior distribution. The estimator fits the switched dynamic equations of the converter, and a whitening filter derived from the model makes the loss the negative log-likelihood of the one-step residuals, whose curvature at the optimum gives the Laplace posterior. Taking one sample per dwell, at its center, keeps the samples away from the switching transitions.

On simulated buck converters, the stated intervals match the scatter of the estimates over repeated noise realizations with three different sampling schemes and two different transient types. At 10~\glspl{LSB}, the median error is \qty{0.33}{\percent}, against \qty{1.9}{\percent} achieved in \cite{Shuai2022} and \qty{3.0}{\percent} for least squares, at about a fifth of the computation time with respect to \cite{Shuai2022}. On a laboratory buck converter every $\pm 2\sigma$ interval overlaps its independent reference, and the posterior of a single record is a good, if slightly conservative, indicator of the variability of the estimates. A reduction of the output capacitance in steps of about \qty{2}{\percent} raises a warning after two to three steps  with no false alarm on the other parameters.

\bibliographystyle{IEEEtran}
\bibliography{IEEEabrv,references}

\begin{thebibliography}{10}
\providecommand{\url}[1]{#1}
\csname url@samestyle\endcsname
\providecommand{\newblock}{\relax}
\providecommand{\bibinfo}[2]{#2}
\providecommand{\BIBentrySTDinterwordspacing}{\spaceskip=0pt\relax}
\providecommand{\BIBentryALTinterwordstretchfactor}{4}
\providecommand{\BIBentryALTinterwordspacing}{\spaceskip=\fontdimen2\font plus
\BIBentryALTinterwordstretchfactor\fontdimen3\font minus
  \fontdimen4\font\relax}
\providecommand{\BIBforeignlanguage}[2]{{%
\expandafter\ifx\csname l@#1\endcsname\relax
\typeout{** WARNING: IEEEtran.bst: No hyphenation pattern has been}%
\typeout{** loaded for the language `#1'. Using the pattern for}%
\typeout{** the default language instead.}%
\else
\language=\csname l@#1\endcsname
\fi
#2}}
\providecommand{\BIBdecl}{\relax}
\BIBdecl

\bibitem{VanderBroeck2023}
C.~H. Van Der~Broeck, S.~Kalker, and R.~W. De~Doncker, ``Intelligent monitoring
  and maintenance technology for next-generation power electronic systems,''
  \emph{IEEE Journal of Emerging and Selected Topics in Power Electronics},
  vol.~11, no.~4, pp. 4403--4418, 2023.

\bibitem{Wang2021}
H.~Wang and F.~Blaabjerg, ``Power electronics reliability: State of the art and
  outlook,'' \emph{IEEE Journal of Emerging and Selected Topics in Power
  Electronics}, vol.~9, no.~6, pp. 6476--6493, 2021.

\bibitem{Miao2024}
J.~Miao, Y.~Liu, Q.~Yin, B.~Ju, G.~Zhang, and H.~Wang, ``A novel soft fault
  detection and diagnosis method for a {DC/DC} buck converter based on
  contrastive learning,'' \emph{IEEE Transactions on Power Electronics},
  vol.~39, no.~1, pp. 1501--1513, 2024.

\bibitem{dc_link_monitoring2}
X.~Zhu, D.~Zhang, A.~Li, and W.~Gao, ``An online noninvasive monitoring
  technique for the {ESR} of an output capacitor in switched-mode power
  supplies,'' \emph{IEEE Transactions on Power Electronics}, vol.~39, no.~11,
  pp. 15\,045--15\,055, 2024.

\bibitem{AlGreer2019}
M.~Al-Greer, M.~Armstrong, M.~Ahmeid, and D.~Giaouris, ``Advances on system
  identification techniques for {DC}–{DC} switch mode power converter
  applications,'' \emph{IEEE Transactions on Power Electronics}, vol.~34,
  no.~7, pp. 6973--6990, 2019.

\bibitem{Lu2023}
Y.~Lu, M.~Zhang, L.~Nordström, and Q.~Xu, ``An online digital twin based
  health monitoring method for boost converter using neural network,'' in
  \emph{2023 IEEE Energy Conversion Congress and Exposition (ECCE)}, 2023, pp.
  3701--3706.

\bibitem{Shuai2022}
S.~Zhao, Y.~Peng, Y.~Zhang, and H.~Wang, ``Parameter estimation of power
  electronic converters with physics-informed machine learning,'' \emph{IEEE
  Transactions on Power Electronics}, vol.~37, no.~10, pp. 11\,567--11\,578,
  2022.

\bibitem{Mann2016}
J.~K. Mann, S.~Perinpanayagam, and I.~Jennions, ``Aging detection capability
  for switch-mode power converters,'' \emph{IEEE Transactions on Industrial
  Electronics}, vol.~63, no.~5, pp. 3216--3227, 2016.

\bibitem{DiNezio2023}
G.~{Di Nezio}, M.~{Di Benedetto}, A.~Lidozzi, and L.~Solero, ``{DC-DC} boost
  converters parameters estimation based on digital twin,'' \emph{IEEE
  Transactions on Industry Applications}, vol.~59, no.~5, pp. 6232--6241, 2023.

\bibitem{DiNezio2022}
G.~Di~Nezio, M.~Di~Benedetto, A.~Lidozzi, and L.~Solero, ``Digital twin based
  real-time analysis of {DC-DC} boost converters,'' in \emph{2022 IEEE Energy
  Conversion Congress and Exposition (ECCE)}, 2022, pp. 1--7.

\bibitem{AlGreer2012}
M.~Algreer, M.~Armstrong, and D.~Giaouris, ``Active online system
  identification of switch mode {DC}–{DC} power converter based on efficient
  recursive {DCD-IIR} adaptive filter,'' \emph{IEEE Transactions on Power
  Electronics}, vol.~27, no.~11, pp. 4425--4435, 2012.

\bibitem{Song2024}
W.~Song, Y.~Zou, C.~Ma, and S.~Zhang, ``Digital twin modeling method of
  three-phase inverter-driven {PMSM} systems for parameter estimation,''
  \emph{IEEE Transactions on Power Electronics}, vol.~39, no.~2, pp.
  2360--2371, 2024.

\bibitem{dc_link_monitoring}
Z.~Zhao, H.~Hu, Z.~He, W.~Lu, H.~H.-C. Iu, F.~Blaabjerg, and P.~Davari, ``A
  transient-modeling-based grey-box method for online monitoring of {DC}-link
  capacitors,'' \emph{IEEE Transactions on Power Electronics}, vol.~38, no.~11,
  pp. 14\,547--14\,562, 2023.

\bibitem{Nazif2025}
P.~Behzad~Nazif, M.~Saeed, S.~Ahmad, J.~M. Guerrero, A.~Rodríguez~Mendez,
  G.~C. Ozaita~Araico, I.~Larrazabal, and F.~Briz, ``Parameter identification
  for {DC-DC} buck converter digital twin considering sensor errors,''
  \emph{IEEE Journal of Emerging and Selected Topics in Power Electronics},
  vol.~13, no.~4, pp. 5111--5123, 2025.

\bibitem{soliman2017artificial}
H.~Soliman, I.~Abdelsalam, H.~Wang, and F.~Blaabjerg, ``Artificial neural
  network based {DC}-link capacitance estimation in a diode-bridge front-end
  inverter system,'' in \emph{2017 IEEE 3rd International Future Energy
  Electronics Conference and ECCE Asia (IFEEC 2017-ECCE Asia)}.\hskip 1em plus
  0.5em minus 0.4em\relax IEEE, 2017, pp. 196--201.

\bibitem{lin2019non}
B.~H. Lin, J.~T. Tsai, and K.~L. Lian, ``A non-invasive method for estimating
  circuit and control parameters of voltage source converters,'' \emph{IEEE
  Transactions on Circuits and Systems I: Regular Papers}, vol.~66, no.~12, pp.
  4911--4921, 2019.

\bibitem{Kalker2022}
S.~Kalker, D.~Meier, C.~H. van~der Broeck, and R.~W. De~Doncker,
  ``Self-calibrating loss models for real-time monitoring of power modules
  based on artificial neural networks,'' in \emph{2022 IEEE Energy Conversion
  Congress and Exposition (ECCE)}, 2022, pp. 1--8.

\bibitem{oruklu2024machine}
K.~{\"O}r{\"u}kl{\"u} and {\c{S}}.~A{\u{g}}alar, ``Machine learning-based
  condition monitoring for {DC}-link capacitors in {AC/DC/AC} converters,''
  \emph{IEEE Transactions on Industrial Electronics}, vol.~72, no.~4, pp.
  4227--4237, 2025.

\bibitem{She2022}
Z.~She and G.~Chen, ``Full-parameter identification of buck converter based on
  time-domain mapping neural network,'' in \emph{2022 IEEE International Power
  Electronics and Application Conference and Exposition (PEAC)}, 2022, pp.
  1000--1004.

\bibitem{Shang2024}
Q.~Shang, F.~Xiao, Y.~Fan, R.~Wang, H.~Qin, and T.~Song, ``Parameter estimation
  of {DAB} converter using intelligent algorithms and steady-state modeling
  considering nonidealities,'' \emph{IEEE Transactions on Industrial
  Electronics}, vol.~71, no.~12, pp. 16\,717--16\,727, 2024.

\bibitem{Dey2025}
S.~Dey and A.~Mallik, ``Physics informed neural network—estimated circuit
  parameter adaptive modulation of {DAB},'' \emph{IEEE Transactions on Power
  Electronics}, vol.~40, no.~10, pp. 14\,821--14\,841, 2025.

\bibitem{wechsler2012condition}
A.~Wechsler, B.~C. Mecrow, D.~J. Atkinson, J.~W. Bennett, and M.~Benarous,
  ``Condition monitoring of {DC}-link capacitors in aerospace drives,''
  \emph{IEEE Transactions on Industry Applications}, vol.~48, no.~6, pp.
  1866--1874, 2012.

\bibitem{de2023real}
S.~{De L{\'o}pez Diz}, R.~{Mart{\'\i}n L{\'o}pez}, F.~J. {Rodr{\'\i}guez
  S{\'a}nchez}, E.~{D{\'\i}az Llerena}, and E.~J. {Bueno Pe{\~n}a}, ``A
  real-time digital twin approach on three-phase power converters applied to
  condition monitoring,'' \emph{Applied Energy}, vol. 334, p. 120606, 2023.

\bibitem{Xie2025}
T.~Xie, Y.~Zhu, X.~Wang, and Y.~Yang, ``Robust data-light parameter estimation
  for {DAB} converters with physics-informed neural network,'' in \emph{2025
  IEEE Energy Conversion Conference Congress and Exposition (ECCE)}, 2025, pp.
  1--7.

\bibitem{heo2013capacitance}
H.-J. Heo, W.-S. Im, J.-S. Kim, and J.-M. Kim, ``A capacitance estimation of
  film capacitors in an {LCL}-filter of grid-connected {PWM} converters,''
  \emph{Journal of Power Electronics}, vol.~13, no.~1, pp. 94--103, 2013.

\bibitem{pirsto2020real}
V.~Pirsto, J.~Kukkola, F.~M.~M. Rahman, and M.~Hinkkanen, ``Real-time
  identification of {LCL} filters employed with grid converters,'' \emph{IEEE
  Transactions on Industry Applications}, vol.~56, no.~5, pp. 5158--5169, 2020.

\bibitem{li2021online}
T.~Li, J.~Chen, P.~Cong, X.~Dai, R.~Qiu, and Z.~Liu, ``Online condition
  monitoring of {DC}-link capacitor for {AC/DC/AC} {PWM} converter,''
  \emph{IEEE Transactions on Power Electronics}, vol.~37, no.~1, pp. 865--878,
  2022.

\bibitem{yao2023estimation}
B.~Yao, Y.~Peng, Y.~Zhang, H.~Wang, and H.~Wang, ``An estimation method of
  high-order {LC} circuits in power electronic converters,'' \emph{IEEE
  Transactions on Industrial Electronics}, vol.~71, no.~5, pp. 5274--5284,
  2024.

\bibitem{Ahmeid2018}
M.~Ahmeid, M.~Armstrong, M.~Al-Greer, and S.~Gadoue, ``Computationally
  efficient self-tuning controller for {DC}–{DC} switch mode power converters
  based on partial update {Kalman} filter,'' \emph{IEEE Transactions on Power
  Electronics}, vol.~33, no.~9, pp. 8081--8090, 2018.

\bibitem{Ahmeid2017}
M.~Ahmeid, M.~Armstrong, S.~Gadoue, M.~Al-Greer, and P.~Missailidis,
  ``Real-time parameter estimation of {DC}–{DC} converters using a self-tuned
  {Kalman} filter,'' \emph{IEEE Transactions on Power Electronics}, vol.~32,
  no.~7, pp. 5666--5674, 2017.

\bibitem{peng2020digital}
Y.~Peng, S.~Zhao, and H.~Wang, ``A digital twin based estimation method for
  health indicators of {DC}--{DC} converters,'' \emph{IEEE Transactions on
  Power Electronics}, vol.~36, no.~2, pp. 2105--2118, 2021.

\bibitem{liu2021condition}
Y.~Liu, G.~Chen, Y.~Liu, L.~Mo, and X.~Qing, ``Condition monitoring of power
  electronics converters based on digital twin,'' in \emph{2021 IEEE 3rd
  International Conference on Circuits and Systems (ICCS)}.\hskip 1em plus
  0.5em minus 0.4em\relax IEEE, 2021, pp. 190--195.

\bibitem{diNezio2023parameters}
G.~Di~Nezio, S.~{De L{\'o}pez Diz}, M.~Di~Benedetto, A.~Lidozzi, E.~J. {Bueno
  Pe{\~n}a}, and L.~Solero, ``Parameters estimation of a 3-phase {AC-DC}
  converter based on the digital twin method,'' in \emph{2023 IEEE Energy
  Conversion Congress and Exposition (ECCE)}.\hskip 1em plus 0.5em minus
  0.4em\relax IEEE, 2023, pp. 2937--2944.

\bibitem{fard2023digital}
M.~T. Fard and J.~He, ``Digital twin health monitoring of five-level {ANPC}
  power converter based on estimation of semiconductor on-state resistance,''
  in \emph{2023 IEEE Industry Applications Society Annual Meeting (IAS)}.\hskip
  1em plus 0.5em minus 0.4em\relax IEEE, 2023, pp. 1--7.

\bibitem{XLi2016}
B.~X. Li and K.~S. Low, ``Low sampling rate online parameters monitoring of
  {DC}–{DC} converters for predictive-maintenance using biogeography-based
  optimization,'' \emph{IEEE Transactions on Power Electronics}, vol.~31,
  no.~4, pp. 2870--2879, 2016.

\bibitem{milton2020controller}
M.~Milton, C.~De~La~O, H.~L. Ginn, and A.~Benigni, ``Controller-embeddable
  probabilistic real-time digital twins for power electronic converter
  diagnostics,'' \emph{IEEE Transactions on Power Electronics}, vol.~35, no.~9,
  pp. 9850--9864, 2020.

\bibitem{zimmer2026parameter}
M.~Zimmer, M.~Milton, D.~Cucak, T.~Pesch, H.~Ginn, and A.~Benigni, ``Parameter
  estimation of power electronic converter by affine physics-informed
  {Gaussian} processes,'' \emph{IEEE Transactions on Power Electronics},
  vol.~41, no.~7, pp. 11\,210--11\,222, 2026.

\bibitem{vanderVaart1998}
A.~W. van~der Vaart, \emph{Asymptotic Statistics}, ser. Cambridge Series in
  Statistical and Probabilistic Mathematics.\hskip 1em plus 0.5em minus
  0.4em\relax Cambridge, U.K.: Cambridge University Press, 1998.

\bibitem{Murphy2022}
\BIBentryALTinterwordspacing
K.~P. Murphy, \emph{Probabilistic Machine Learning: An Introduction}.\hskip 1em
  plus 0.5em minus 0.4em\relax MIT Press, 2022. [Online]. Available:
  \url{https://probml.github.io/pml-book/book1.html}
\BIBentrySTDinterwordspacing

\bibitem{Hairer1993}
E.~Hairer, S.~P. N{\o}rsett, and G.~Wanner, \emph{Solving Ordinary Differential
  Equations {I}: Nonstiff Problems}, 2nd~ed., ser. Springer Series in
  Computational Mathematics.\hskip 1em plus 0.5em minus 0.4em\relax Berlin,
  Germany: Springer, 1993, vol.~8.

\bibitem{Buso2015}
S.~Buso and P.~Mattavelli, \emph{Digital Control in Power Electronics},
  2nd~ed., ser. Synthesis Lectures on Power Electronics.\hskip 1em plus 0.5em
  minus 0.4em\relax San Rafael, CA, USA: Morgan \& Claypool, 2015.

\bibitem{corradini2008multisampled}
L.~Corradini, P.~Mattavelli, E.~Tedeschi, and D.~Trevisan, ``High-bandwidth
  multisampled digitally controlled {DC--DC} converters using ripple
  compensation,'' \emph{IEEE Transactions on Industrial Electronics}, vol.~55,
  no.~4, pp. 1501--1508, 2008.

\bibitem{lukic2005sigma}
Z.~Luki{\'c}, K.~Wang, and A.~Prodi{\'c}, ``High-frequency digital controller
  for dc-dc converters based on multi-bit {$\Sigma$-$\Delta$} pulse-width
  modulation,'' in \emph{Proc. IEEE Applied Power Electronics Conference and
  Exposition (APEC)}, vol.~1, 2005, pp. 35--40.

\bibitem{Golub2013}
G.~H. Golub and C.~F. {Van Loan}, \emph{Matrix Computations}, 4th~ed.\hskip 1em
  plus 0.5em minus 0.4em\relax Baltimore, MD, USA: Johns Hopkins University
  Press, 2013, secs.~4.1.2 and 4.5.

\bibitem{Lutkepohl2005}
H.~L{\"u}tkepohl, \emph{New Introduction to Multiple Time Series
  Analysis}.\hskip 1em plus 0.5em minus 0.4em\relax Berlin, Germany: Springer,
  2005.

\bibitem{Brockwell1991}
P.~J. Brockwell and R.~A. Davis, \emph{Time Series: Theory and Methods},
  2nd~ed.\hskip 1em plus 0.5em minus 0.4em\relax New York, NY, USA: Springer,
  1991.

\bibitem{Kingma2015}
D.~P. Kingma and J.~Ba, ``Adam: A method for stochastic optimization,'' in
  \emph{Proc. 3rd Int. Conf. Learning Representations (ICLR)}, San Diego, CA,
  USA, 2015.

\bibitem{Nocedal2006}
J.~Nocedal and S.~J. Wright, \emph{Numerical Optimization}, 2nd~ed.\hskip 1em
  plus 0.5em minus 0.4em\relax New York, NY, USA: Springer, 2006.

\bibitem{Marquardt1963}
D.~W. Marquardt, ``An algorithm for least-squares estimation of nonlinear
  parameters,'' \emph{Journal of the Society for Industrial and Applied
  Mathematics}, vol.~11, no.~2, pp. 431--441, 1963.

\bibitem{Transtrum2010}
M.~K. Transtrum, B.~B. Machta, and J.~P. Sethna, ``Why are nonlinear fits to
  data so challenging?'' \emph{Physical Review Letters}, vol. 104, no.~6, p.
  060201, 2010.

\bibitem{Transtrum2011}
------, ``Geometry of nonlinear least squares with applications to sloppy
  models and optimization,'' \emph{Physical Review E}, vol.~83, no.~3, p.
  036701, 2011.

\bibitem{Liu1989}
D.~C. Liu and J.~Nocedal, ``On the limited memory {BFGS} method for large scale
  optimization,'' \emph{Mathematical Programming}, vol.~45, no. 1--3, pp.
  503--528, 1989.

\bibitem{ti_tms320_datasheet}
\BIBentryALTinterwordspacing
\emph{TMS320F2837xD Dual-Core {Delfino} Microcontroller}, Texas Instruments,
  Dallas, TX, USA, Dec. 2013. [Online]. Available:
  \url{https://www.ti.com/lit/ds/symlink/tms320f28379d.pdf}
\BIBentrySTDinterwordspacing

\bibitem{Holm1979}
S.~Holm, ``A simple sequentially rejective multiple test procedure,''
  \emph{Scandinavian Journal of Statistics}, vol.~6, no.~2, pp. 65--70, 1979.

\bibitem{mosfet_datasheet}
\BIBentryALTinterwordspacing
\emph{{IMZ120R045M1} {CoolSiC}\texttrademark{} 1200\,V SiC Trench {MOSFET}},
  Infineon Technologies, Neubiberg, Germany, Dec. 2020. [Online]. Available:
  \url{https://www.infineon.com/assets/row/public/documents/60/49/infineon-imz120r045m1-datasheet-en.pdf}
\BIBentrySTDinterwordspacing

\bibitem{kennedy2001bayesian}
M.~C. Kennedy and A.~O'Hagan, ``{Bayesian} calibration of computer models,''
  \emph{Journal of the Royal Statistical Society: Series B (Statistical
  Methodology)}, vol.~63, no.~3, pp. 425--464, 2001.

\bibitem{inductor_datasheet}
\BIBentryALTinterwordspacing
\emph{{HCTI} Series}, Bel Fuse Inc. (Signal Transformer), Lynbrook, NY, USA,
  2019. [Online]. Available:
  \url{https://www.belfuse.com/media/datasheets/products/chokes-coils-inductors/ds-st-high-current-torodial-inductors-series.pdf}
\BIBentrySTDinterwordspacing

\bibitem{imaoka2019powder}
J.~Imaoka, K.~Okamoto, M.~Shoyama, Y.~Ishikura, M.~Noah, and M.~Yamamoto,
  ``Modeling, magnetic design, simulation methods, and experimental evaluation
  of various powder cores used in power converters considering their {DC}
  superimposition characteristics,'' \emph{IEEE Transactions on Power
  Electronics}, vol.~34, no.~9, pp. 9033--9051, 2019.

\bibitem{rayleigh1887}
{Lord Rayleigh}, ``Notes on electricity and magnetism.---{III}. {O}n the
  behaviour of iron and steel under the operation of feeble magnetic forces,''
  \emph{The London, Edinburgh, and Dublin Philosophical Magazine and Journal of
  Science}, vol.~23, no. 142, pp. 225--245, 1887.

\bibitem{preisach2017aftereffect}
F.~Preisach, ``On the magnetic aftereffect,'' \emph{IEEE Transactions on
  Magnetics}, vol.~53, no.~3, pp. 1--11, 2017, english translation of
  ``{\"U}ber die magnetische {N}achwirkung,'' Zeitschrift f{\"u}r Physik, 1935.

\bibitem{small_c_datasheet}
\BIBentryALTinterwordspacing
\emph{{EGT335M1HD11RR}}, Samxon. [Online]. Available:
  \url{https://www.lcsc.com/product-detail/aluminum-electrolytic-capacitors-leaded_man-yue-tech-egt335m1hd11rr_C913116.html}
\BIBentrySTDinterwordspacing

\bibitem{Kong2026}
J.~Kong, D.~Zhou, X.~Wei, S.~Zhao, Z.~Yuan, Y.~Xu, and H.~Wang,
  ``Physics-informed neural network based noninvasive lc-filter parameter
  estimation in inverters,'' \emph{IEEE Transactions on Industrial
  Electronics}, pp. 1--13, 2026.

\end{thebibliography}
\begin{IEEEbiography}[{\includegraphics[width=1in,height=1.25in,clip,keepaspectratio]{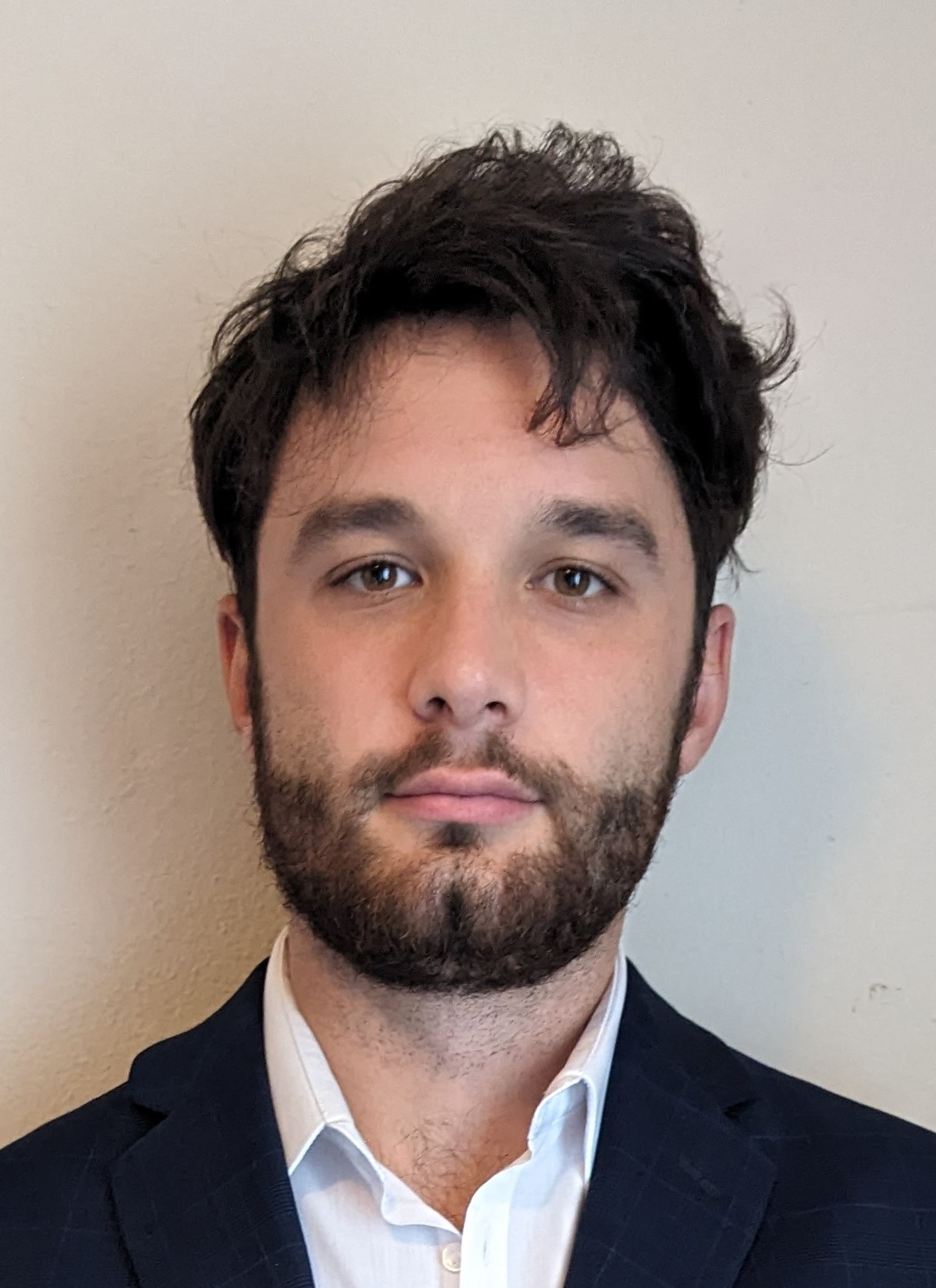}}]
    {Tomas Monopoli} received a double M.Sc. degree (cum laude) from the University of Politecnico di Milano and Politecnico di Torino, Italy, in 2020, and in 2025 he received the Ph.D. in Electrical Engineering from Politecnico di Milano.
    He is currently a postdoctoral researcher with the Department of Energy at Aalborg University, Denmark.
    In 2021 he was with the European Space Agency (ESA), ESA/ESTEC, The Netherlands, as a Research Fellow. His research interests include power electronic reliability, condition and health monitoring of power converters, statistical and physics-informed machine learning for power electronics and electromagnetic compatibility.
\end{IEEEbiography}

\begin{IEEEbiography}[{\includegraphics[width=1in,height=1.25in,clip,keepaspectratio]{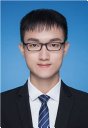}}]
    {Jiahong Liu} (Member, IEEE) received the B.E. degree in electrical engineering and automation and the M.E. degree in electrical engineering from Wuhan University, Wuhan, China, in 2019 and 2021, respectively, and the Ph.D. degree in electrical engineering from Aalborg University, Aalborg, Denmark, in 2025.
    From November to December 2024, he was a Visiting Scholar with Corporate Research Center, ABB, Sweden. He is a Postdoctoral Researcher with the Department of Energy, Aalborg University. His research interests include power electronic reliability, fault protection, condition and health monitoring for power semiconductor devices and converters, and solid-state transformers.
\end{IEEEbiography}

\begin{IEEEbiography}[{\includegraphics[width=1in,height=1.25in,clip,keepaspectratio]{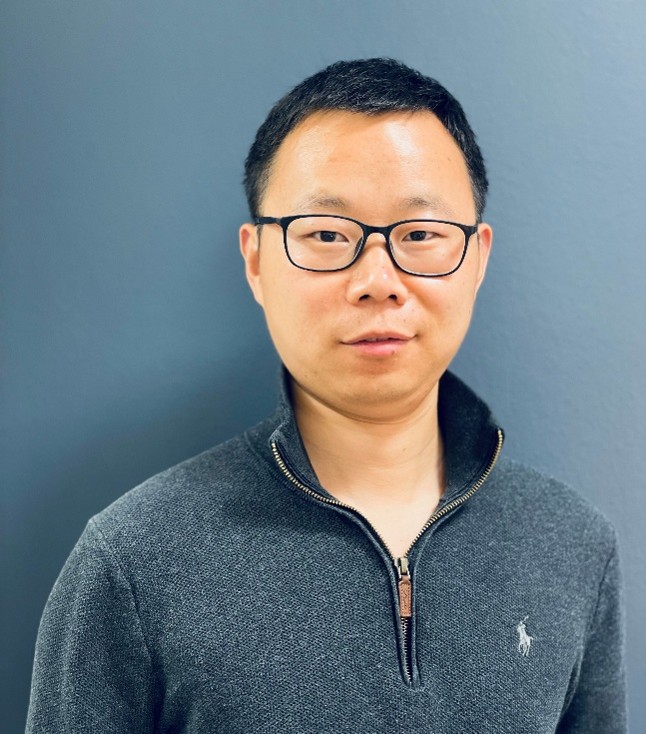}}]{Shuai Zhao}
    (Senior Member, IEEE) received the B.S., M.S., and Ph.D. degrees in information and telecommunication engineering from Northwestern Polytechnical University, Xi’an, China, in 2011, 2014, and 2018, respectively. He is currently an Assistant Professor with AAU Energy, Aalborg University, Denmark, where he was a Postdoc from 2018 to 2022. He was a visiting Ph.D. student with University of Toronto, Canada, from 2014 to 2016, and a visiting scholar with University of Texas at Dallas, U.S., in 2018. He received the Top Paper Award of IEEE Power Electronics Magazine in 2025 and the Prize Paper Award (Second Place) of IEEE Transactions on Power Electronics in 2024. He currently serves as an Associate Editor of IEEE JOURNAL OF EMERGING AND SELECTED TOPICS IN POWER ELECTRONICS, IEEE TRANSACTIONS ON INDUSTRY APPLICATIONS, and IEEE TRANSACTIONS ON VEHICLE TECHNOLOGY. His research interests include physics-informed machine learning, system informatics, condition monitoring, diagnostics and prognostics, and tailored AI tools for power electronic systems.
\end{IEEEbiography}

\vfill

\end{document}